\documentclass{article}
\pdfoutput=1
\usepackage[utf8]{inputenc}
\usepackage{graphicx,psfrag,epsf}
\usepackage{enumerate}
\usepackage{natbib}
\usepackage{mathtools}
\usepackage{booktabs}
\usepackage{color}
\usepackage[english]{babel} 
\usepackage[protrusion=true,expansion=true]{microtype} 
\usepackage{amsmath,amsfonts,amsthm,amssymb}
\usepackage{dsfont}
\usepackage{subfigure}
\usepackage{chngcntr}
\usepackage[toc,page]{appendix}
\usepackage{textcomp}
\usepackage{url}

\usepackage{array}
\usepackage{booktabs} 
\usepackage{natbib}
\usepackage{setspace}
\usepackage{amsmath}

\usepackage{soul}
\usepackage{multirow}
\usepackage{enumitem}
\usepackage{microtype} 
\usepackage{xcolor}
\usepackage{tabularx}
\usepackage[font = small,labelfont=bf,textfont=it]{caption} 
\usepackage{footnote}
\usepackage{algorithm}
\usepackage[ruled, vlined,algo2e]{algorithm2e}
\usepackage[noend]{algpseudocode}
\usepackage{etoolbox}
\usepackage{tabularx}
\usepackage{authblk}
\usepackage{caption}
\usepackage{indentfirst}
\usepackage{enumitem}
\newlist{steps}{enumerate}{1}
\setlist[steps, 1]{label = Step \arabic*:}
\usepackage[colorlinks,citecolor=blue,urlcolor=blue,filecolor=blue,backref=page]{hyperref}

\algblock{ParFor}{EndParFor}
\algnewcommand\algorithmicparfor{\textbf{for}}
\algnewcommand\algorithmicpardo{\textbf{do\ parallel}}
\algnewcommand\algorithmicendparfor{\textbf{end\ parallel\ for}}
\algrenewtext{ParFor}[1]{\algorithmicparfor\ #1\ \algorithmicpardo}
\algrenewtext{EndParFor}{\algorithmicendparfor}

\makeatletter
\def\BState{\State\hskip-\ALG@thistlm}
\DeclareMathOperator*{\argmax}{\arg\!\max}
\newcommand{\distas}[1]{\mathbin{\overset{#1}{\kern\z@\sim}}}%
\newcommand{\bm}[1]{\mathbf{#1}}
\newsavebox{\mybox}\newsavebox{\mysim}
\newcommand{\distras}[1]{%
  \savebox{\mybox}{\hbox{\kern3pt$\scriptstyle#1$\kern3pt}}%
  \savebox{\mysim}{\hbox{$\sim$}}%
  \mathbin{\overset{#1}{\kern\z@\resizebox{\wd\mybox}{\ht\mysim}{$\sim$}}}%
}
\newtheorem{theorem}{Theorem}

\newtheorem{proposition}[theorem]{Proposition}
\newtheorem{corollary}{Corollary}

\newcommand{\be}{\begin{equation}}
\newcommand{\ee}{\end{equation}}
\newcommand{\bi}{\begin{itemize}}
\newcommand{\ei}{\end{itemize}}
\newcommand{\ben}{\begin{enumerate}}
\newcommand{\een}{\end{enumerate}}

\newcolumntype{K}[1]{\geq {\centering\arraybackslash}p{#1}}
\allowdisplaybreaks

\makeatother

\let\oldbibliography\thebibliography
\renewcommand{\thebibliography}[1]{\oldbibliography{#1}
\setlength{\itemsep}{0pt}} 

\patchcmd{\footnotemark}{\stepcounter{footnote}}{\refstepcounter{footnote}}{}{}

\newtheorem{lemma}[theorem]{Lemma}
\newtheorem{definition}[theorem]{Definition}

\title{Bayesian Uncertainty Quantification for\\ Low-Rank Matrix Completion}
\author{Henry Shaowu Yuchi\footnote{H. Milton Stewart School of Industrial \& Systems Engineering, Georgia Institute of Technology} $^{\ddagger}$\quad Simon Mak\footnote{Department of Statistical Science, Duke University} \footnote{Joint first authors.}\quad Yao Xie$^{*}$}
\date{November 2021}

\begin{document}

\maketitle
\begin{abstract}
	We consider the problem of uncertainty quantification for an unknown low-rank matrix $\bm{X}$, given a partial and noisy observation of its entries. This quantification of uncertainty is essential for many real-world problems, including image processing, satellite imaging, and seismology, providing a principled framework for validating scientific conclusions and guiding decision-making. However, existing literature has mainly focused on the completion (i.e., point estimation) of the matrix $\bm{X}$, with little work on investigating its uncertainty. To this end, we propose in this work a new Bayesian modeling framework, called BayeSMG, which parametrizes the unknown $\bm{X}$ via its underlying row and column subspaces. This Bayesian subspace parametrization enables efficient posterior inference on matrix subspaces, which represents interpretable phenomena in many applications. This can then be leveraged for improved matrix recovery. We demonstrate the effectiveness of BayeSMG over existing Bayesian matrix recovery methods in numerical experiments, image inpainting, and a seismic sensor network application. 
\end{abstract}

\vfill

\section{Introduction}

Low-rank matrices play a vital role in modeling many scientific and engineering problems, including (but not limited to) image processing, satellite imaging, and network analysis. In such applications, however, only a small portion of the desired matrix (which we denote as $\bm{X} \in \mathbb{R}^{m_1 \times m_2}$ in this article) can be observed. The reasons for this are two-fold: (i) the cost of observing all matrix entries can be high, requiring expensive computational, experimental, or communication expenditure; (ii) there can be missing observations at individual entries due to sensor malfunction, experimental failure, or unreliable data transmission. The \textit{matrix completion} problem aims to complete the missing entries of $\bm{X}$ from a partial (and often-times noisy) observation. Matrix completion has attracted much attention since the seminal works of \cite{CT2010}, \cite{CR2012}, and \cite{Rec2011}. The theory and methodology behind point estimation are now well-understood for matrix completion, under the assumption that $\bm{X}$ is low-rank, with various convex and non-convex optimization algorithms developed for performing this recovery. 

However, much of the literature (a detailed review is in Section \ref{sec:lit}) has focused on the completion, i.e., \textit{point estimation}, of $\bm{X}$, with little work on exploring the uncertainty of such estimates. In many scientific and engineering applications, such estimates are much more useful when coupled with a measure of uncertainty. The principled characterization (and reduction) of this uncertainty is known as \textit{uncertainty quantification} (UQ), see, e.g., \cite{Smi2013}. UQ is becoming increasingly important in various applications, providing a principled framework for validating scientific conclusions and guiding decision-making. 

In this paper, we address the problem of UQ for the matrix completion problem from a Bayesian perspective. We propose a novel Bayesian modeling framework, called BayeSMG, which quantifies uncertainty in the desired matrix $\bm{X}$ via posterior sampling on its underlying subspaces. BayeSMG can be viewed as a hierarchical Bayesian extension of the singular matrix-variate Gaussian (SMG) distribution (see \citealp{GN1999,MX2017}), with hierarchical priors on matrix subspaces. A scalable posterior sampling algorithm is then derived for BayeSMG, which leverages the efficient subspace sampling algorithms proposed in \cite{hoff2007model} and \cite{Hof2009}. By integrating the subspace structure for posterior inference, we show that BayeSMG enjoys improved recovery performance and better interpretability compared with existing Bayesian models in extensive numerical experiments and a real-world seismic sensor network application. 
\subsection{Existing literature}
\label{sec:lit}
Much of the existing literature on inferring $\bm{X}$ from partial observations falls under the topic of \textit{matrix completion} - the completion (or point estimation) of $\bm{X}$ from observed entries. Early works in this area include the seminal works of \cite{CT2010}, \cite{CR2012}, and \cite{Rec2011}, which established conditions for exact completion via nuclear-norm minimization, under the assumption that observations are uniformly sampled without noise. This is then extended to the \textit{noisy} matrix completion setting, where entries are observed with noise; important results include \cite{CP2010}, \cite{Rea2010}, \cite{Kea2011}, and \cite{NW2012}, among others. There is now a rich body of work on matrix completion; recent overviews include \cite{DR2016} and \cite{chi2019nonconvex}. However, completion focuses solely on the point estimation of matrix entries and does not provide uncertainty quantification on those unobserved. In scenarios where only a few entries are observed(see motivating applications), this uncertainty can be as valuable as point estimates in assessing the quality of the recovered matrix.

The current research literature has generally focused on point estimation of the unknown matrix $\bm{X}$. The problem of quantifying uncertainties in $\bm{X}$ has been relatively unexplored, but it is nonetheless an important one given the motivating applications. One recent pioneering work on this is \cite{chen2019inference}, which proposed entrywise confidence intervals for both convex and non-convex estimators on $\bm{X}$, via debiasing using low-rank factors of the matrix. The resulting debiased estimators admit nearly precise nonasymptotic distributional characterizations, which in turn enable optimal construction of confidence intervals for missing matrix entries and low-rank factors. Our approach has several distinctions from this work. First, the latter is a frequentist approach with appealing theoretical guarantees, whereas our approach is Bayesian and yields a richer quantification of uncertainty on $\bm{X}$ via a hierarchical Bayesian model. Second, to derive elegant theoretical results, the latter requires a sample size complexity condition on $\bm{X}$, similar to the minimum sample size condition in standard matrix completion analysis (see, e.g., \citealp{CR2012}). Our UQ approach, in contrast, is applicable for any sample size $n$ on $\bm{X}$, particularly for the ``small-$n$'' setting where observations are limited and uncertainty quantification is most needed.

Another approach for quantifying uncertainty is via Bayesian modeling. There is a growing literature on Bayesian matrix completion, of which the most popular approach is the Bayesian Probabilistic Matrix Factorization (BPMF) method in \cite{SM2008}. BPMF adopts the following probabilistic model on $\bm{X}$:
$\bm{X} = \bm{M} \bm{N}^T$, $\bm{M} \in \mathbb{R}^{m_1 \times R}$, $\bm{N} \in \mathbb{R}^{m_2 \times R},$
where $R < m_1 \wedge m_2 := \min (m_1, m_2)$ is an upper bound on matrix rank. Each row of the factorized matrices $\bm{M}$ and $\bm{N}$ are then assigned i.i.d. Gaussian priors $\mathcal{N}(\boldsymbol{\mu}_M,\boldsymbol{\Sigma_M})$ and $\mathcal{N}(\boldsymbol{\mu}_N,\boldsymbol{\Sigma_N})$, respectively. Conjugate normal hyperpriors are then assigned on the row and column means $\boldsymbol{\mu}_M\sim\mathcal{N}(\bm{0},\boldsymbol{\Sigma}_M\boldsymbol{\beta})$, $\boldsymbol{\mu}_N \sim\mathcal{N}(\bm{0},\boldsymbol{\Sigma_N}\boldsymbol{\beta})$, with Inverse-Wishart hyperpriors on row and column covariance matrices $\boldsymbol{\Sigma}_M \sim \text{IW}(R,\bm{W}), \boldsymbol{\Sigma}_N \sim \text{IW}(R,\bm{W})$. The hyperparameters $\boldsymbol{\beta}$ and $\bm{W}$ are typically specified to provide weakly- or non-informative priors. This model allows for an efficient Gibbs sampler, which performs conjugate sampling on each \textit{row} of $\bm{M}$ and each \textit{row} of $\bm{N}$, along with conjugate updates on the mean vectors $(\boldsymbol{\mu}_M,\boldsymbol{\mu}_N)$ and covariance matrices $(\boldsymbol{\Sigma}_M,\boldsymbol{\Sigma}_N)$. With this, the BPMF can be shown to tackle problems as large as the Netflix dataset, with millions of user-movie ratings. A similar Bayesian model was proposed in \cite{MA2015}, with priors on each entry of $\bm{M}$ and $\bm{N}$. Many other existing Bayesian matrix completion methods (e.g., \citealp{LU2009,Zea2010,Bea2011,Aea2014}) can be viewed as variations or extensions of this BPMF framework.

From a modeling perspective, the key novelty in BayeSMG model is that it requires orthonormality in the factorized matrices, whereas the BPMF does not. Such a factorization can be viewed as parametrizing $\bm{X}$ via its singular value decomposition (SVD). This yields several advantages for our method, which we demonstrate later. First, by explicitly parametrizing row and column subspaces as model parameters, BayeSMG can incorporate prior knowledge on subspaces within the prior specification of such parameters. This prior information is often available in many signal processing and image processing problems, e.g., known signal structure or image features. Second, BayeSMG allows for \textit{direct} inference on subspaces of $\bm{X}$ via posterior sampling, which is of direct interest in many problems, e.g., in sensor network localization (\citealp{zhang2020matrix}; an application we tackle later on) and topology identification problems \citep{eriksson2012high}. For subspace inference, our approach avoids performing an additional SVD step for every posterior sample (compared to the BPMF), which significantly speeds up inference for high-dimensional problems. Finally and perhaps most importantly, BayeSMG can leverage this posterior learning on subspaces to provide improved inference on $\bm{X}$. Compared to the BPMF, our approach can yield faster posterior contraction for unobserved entries when the underlying matrix has a low-rank structure, in both numerical simulations and applications. It enables a more accurate estimate and more precise uncertainty quantification of $\bm{X}$ over the BPMF.

The BayeSMG model also provides several novel theoretical insights. In Section \ref{sec:uq}, we show that the maximum a posteriori (MAP) estimator takes the form of a regularized matrix estimator, which provides a connection between the proposed method and existing matrix completion techniques. We also show that the BayeSMG model provides a probabilistic model on matrix {coherence} \citep{CR2012}. Coherence has been widely used in the matrix completion literature as a theoretical condition for recovery, which measures the ``recoverability'' of a low-rank matrix. Through this, we then establish an error monotonicity result for BayeSMG, which provides a reassuring check on the UQ performance of the proposed model.

The paper is organized as follows. Section \ref{sec:mod} introduces the BayeSMG model. Section \ref{sec:prior} presents an efficient posterior sampling algorithm for $\bm{X}$ via manifold sampling on its subspaces. Section \ref{sec:uq} reveals connections between the BayeSMG model and coherence, and its impact on error convergence. Section \ref{sec:numerical} investigates numerical experiments with synthetic and image data. Section \ref{sec:seismic} explores a real-world seismic sensor network application. Section \ref{sec:conclusion} concludes with discussions.
\section{The SMG model}
\label{sec:mod}
We first describe the Singular Matrix-variate Gaussian (SMG) distribution, and how it can be utilized for modeling matrix subspaces.
\subsection{Problem set-up}

Let $\bm{X} \in \mathbb{R}^{m_1 \times m_2}$ be the matrix of interest, and assume $\bm{X}$ is low-rank, i.e., $R := \text{rank}(\bm{X}) \ll m_1 \wedge m_2$. Let $[m] := \{1, \cdots, m\}$. Suppose $\bm{X}$ is sampled with noise at an index set $\Omega \subseteq [m_1] \times [m_2]$ of size $|\Omega| = n$, yielding observations:
\begin{equation}
Y_{i,j} = X_{i,j} + \epsilon_{i,j}, \quad (i,j) \in \Omega. \label{eq:obs}
\end{equation}
Here, $Y_{i,j}$ is the observation at entry indexed by $(i,j)$, corrupted by noise $\epsilon_{i,j}$. In this work, we assume $\epsilon_{i,j} \distas{i.i.d.} \mathcal{N}(0,\eta^2)$, i.e., the noise on each entry follows an i.i.d. Gaussian distribution with zero mean and variance $\eta^2$. Furthermore, let $\bm{Y}_{\Omega} := (Y_{i,j})_{(i,j) \in \Omega} \in \mathbb{R}^n$ denote the vector of noisy observations, and let $\bm{X}_{\Omega^c}$ be the vector of unobserved matrix entries, where $\Omega^c := ([m_1] \times [m_2]) \setminus \Omega$ is the set of unobserved indices.

With this framework, the desired goal of uncertainty quantification (UQ) can be made more concrete. Given noisy observations $\bm{Y}_{\Omega}$, we wish to not only estimate
the unobserved matrix entries $\bm{X}_{\Omega^c}$, but also quantify a notion of \textit{uncertainty} on both observed or unobserved entries (since observation noise is present).
\subsection{SMG model}

We adopt the following SMG model for the low-rank matrix $\bm{X}$, which we assume to be normal with a zero mean. 
\begin{definition}[SMG model, 
Definition 2.4.1 of \citealp{GN1999}] Let $\bm{Z} \in \mathbb{R}^{m_1 \times m_2}$ be a random matrix with entries $Z_{i,j} \distas{i.i.d.} \mathcal{N}(0, \sigma^2)$ for $(i,j) \in [m_1] \times [m_2]$. The random matrix $\bm{X}$ has a \textup{singular matrix-variate Gaussian (SMG)} distribution if $\bm{X} \stackrel{d}{=} \mathcal{P}_{\mathcal{U}} \bm{Z} \mathcal{P}_{\mathcal{V}}$ for some choice of projection matrices $\mathcal{P}_{\mathcal{U}} = \bm{U}\bm{U}^T$ and $\mathcal{P}_{\mathcal{V}} = \bm{V}\bm{V}^T$, where $\bm{U} \in \mathbb{R}^{m_1 \times R}$, $\bm{U}^T\bm{U} = \bm{I}$,  $\bm{V} \in \mathbb{R}^{m_2 \times R}$, $\bm{V}^T\bm{V} = \bm{I}$ and $R < m_1 \wedge m_2$. We will denote this as $\bm{X} \sim \mathcal{SMG}(\mathcal{P}_{\mathcal{U}},\mathcal{P}_{\mathcal{V}},\sigma^2,R)$.
\label{def:smg}
\end{definition}
In other words, a realization from the SMG distribution can be obtained by first (i) simulating a matrix $\bm{Z}$ from a Gaussian ensemble with variance $\sigma^2$, i.e., a matrix with i.i.d. $\mathcal{N}(0,\sigma^2)$ entries, then (ii) performing a left and right projection of $\bm{Z}$ using the projection matrices $\mathcal{P}_{\mathcal{U}}$ and $\mathcal{P}_{\mathcal{V}}$. Recall that the projection operator $\mathcal{P}_{\mathcal{U}} = \bm{U}\bm{U}^T \in \mathbb{R}^{m_1 \times m_1}$ maps a vector in $\mathbb{R}^{m_1}$ to its orthogonal projection on the $R$-dimensional subspace $\mathcal{U}$ spanned by the columns of $\bm{U}$. By performing this projection, the resulting matrix $\bm{X}=\mathcal{P}_{\mathcal{U}}\bm{Z} \mathcal{P}_{\mathcal{V}}$ can be shown to be of rank $R < m_1 \wedge m_2$, with its row and column spaces $\mathcal{U}$ and $\mathcal{V}$ corresponding to the subspaces for $\mathcal{P}_{\mathcal{U}}$ and $\mathcal{P}_{\mathcal{V}}$. The matrix $\bm{X}$ also lies in the space $\mathcal{T}: =\bigcup_{u_k\in\mathcal{U},v_k\in\mathcal{V}}\text{span}(\{ \bm{u}_k \bm{v}_k^T \}_{k=1}^R)$. With a small choice of $R$, this provides a flexible probabilistic model for the low-rank matrix $\bm{X}$. 

The SMG distribution provides several appealing properties for modeling low-rank matrices. First, it provides a prior modeling framework on the matrix $\bm{X}$ involving its row and column subspaces $\mathcal{U}$ and $\mathcal{V}$. It is known from \cite{Chi2012} that, for each projection operator $\mathcal{P} \in \mathbb{R}^{m \times m}$ of rank $R$, there exists a unique $R$-dimensional hyperplane (or an $R$-plane) in $\mathbb{R}^m$ containing the origin which corresponds to the image of such a projection. It connects the space of rank $R$ projection matrices and the \textit{Grassmann manifold} $\mathcal{G}_{R,m-R}$, the space of $R$-planes in $\mathbb{R}^m$. Viewed this way, the projection matrices parametrizing $\bm{X} \sim \mathcal{SMG}(\mathcal{P}_{\mathcal{U}},\mathcal{P}_{\mathcal{V}},\sigma^2,R)$ encode useful information on the row and column spaces of $\bm{X}$. Second, since the projection of a Gaussian random vector is still Gaussian, the left-right projection of the Gaussian ensemble $\bm{Z}$ results in each entry of $\bm{X}$ being Gaussian-distributed as well. It is useful for deriving a UQ property of the BayeSMG model. 

We now show several distributional properties of the SMG model:

\begin{lemma}[Distributional properties of SMG]
Let $\bm{X}\sim \mathcal{SMG} (\mathcal{P_U},\mathcal{P_V},\sigma^2,R)$, with $\mathcal{P_U}\in\mathbb{R}^{m_1\times m_1}$, $\mathcal{P_V}\in\mathbb{R}^{m_2\times m_2}$, $\sigma^2>0$ and $R< m_1 \wedge m_2$ known. Then:
\begin{enumerate}
    \item[(a)] The density of $\bm{X}$ is given by
    \begin{equation}
        p(\bm{X})=(2\pi\sigma^2)^{-R^2/2} {\rm etr}\left\{ -\frac{1}{2\sigma^2}[(\bm{X}\mathcal{P_V})^T(\mathcal{P_U}\bm{X})]\right\}, \quad \bm{X}\in\mathcal{T},
    \end{equation}
    where ${\rm etr}(\cdot):=\exp\{{\rm tr}(\cdot)\}$.
    \item[(b)] Consider the block decomposition of $\mathcal{P_V}\otimes \mathcal{P_U}$:
    \begin{equation}
        \mathcal{P_V}\otimes \mathcal{P_U}=
        \begin{pmatrix}
        (\mathcal{P_V}\otimes \mathcal{P_U})_\Omega & (\mathcal{P_V}\otimes \mathcal{P_U})_{\Omega,\Omega^c}\\
        (\mathcal{P_V}\otimes \mathcal{P_U})^T_{\Omega,\Omega^c} & (\mathcal{P_V}\otimes \mathcal{P_U})_{\Omega^c}
        \end{pmatrix}.
    \end{equation}
    Conditional on the observed noisy entries $\bm{Y}_\Omega$, the unobserved entries $\bm{X}_{\Omega^c}$ follow the distribution, $    [\bm{X}_{\Omega^c}|\bm{Y}_{\Omega}]\sim \mathcal{N}(\bm{X}_{\Omega^c}^P,\boldsymbol{\Sigma}_{\Omega^c}^P)$. Here, $\gamma^2=\eta^2/\sigma^2$, and
    \begin{equation}
        \begin{split}
    \bm{R}_N(\Omega)&:= (\mathcal{P_V}\otimes \mathcal{P_U})_\Omega\in\mathbb{R}^{N\times N},\\
        \bm{X}_{\Omega^c}^P&:= (\mathcal{P_V}\otimes \mathcal{P_U})^T_{\Omega,\Omega^c}[\bm{R}_N(\Omega)+\gamma^2\bm{I}]^{-1}\bm{Y}_\Omega,\\
        \boldsymbol{\Sigma}_{\Omega^c}^P&:= \sigma^2 \{ (\mathcal{P_V}\otimes \mathcal{P_U})_{\Omega^c}- (\mathcal{P_V}\otimes \mathcal{P_U})^T_{\Omega,\Omega^c}[\bm{R}_N(\Omega)+\gamma^2\bm{I}]^{-1}(\mathcal{P_V}\otimes \mathcal{P_U})^T_{\Omega,\Omega^c}\}.
        \label{eq:condp}
        \end{split}
    \end{equation}
    \item[(c)] Conditional on the observed noisy entries $\bm{Y}_\Omega$, the corresponding entries in $\bm{X}$, namely $\bm{X}_\Omega$, follow the distribution $    [\bm{X}_{\Omega}|\bm{Y}_\Omega]\sim\mathcal{N}(\bm{X}_\Omega^P, \boldsymbol{\Sigma}_\Omega^P)$, where $\otimes$ is the Kronecker product, and
    \begin{equation}
    \begin{split}
    \bm{X}_\Omega^P &:= (\mathcal{P_V}\otimes \mathcal{P_U})_\Omega[\bm{R}_N(\Omega)+\gamma^2\bm{I}]^{-1}\bm{Y}_\Omega,\\
    \boldsymbol{\Sigma}_\Omega^P& := \sigma^2 \{  (\mathcal{P_V}\otimes \mathcal{P_U})_\Omega-(\mathcal{P_V}\otimes \mathcal{P_U})^T_\Omega [\bm{R}_N(\Omega)+\gamma^2\bm{I}]^{-1}(\mathcal{P_V}\otimes \mathcal{P_U})_\Omega\}.
    \label{eq:condp2}
    \end{split}
    \end{equation}
\end{enumerate}
\label{thm:smg}
\end{lemma} 
\noindent {\it Remark:} Lemma \ref{thm:smg} reveals two key properties of the SMG model. First, \textit{prior} to observing data, part (a) shows that the low-rank matrix $\bm{X}$ lies on the space $\mathcal{T}$, and follows a degenerate multivariate Gaussian distribution with mean zero and covariance matrix $\sigma^2 (\mathcal{P}_{\mathcal{V}} \otimes \mathcal{P}_{\mathcal{U}})$. Second, \textit{after} observing the noisy entries $\bm{Y}_{\Omega}$, part (b) shows that the conditional distribution of $\bm{X}_{\Omega^c}$ (the unobserved entries in $\bm{X}$) given $\bm{Y}_{\Omega}$ is still multivariate Gaussian, with closed-form expressions for its mean vector $\bm{X}_{\Omega^c}^P$ and covariance matrix $\boldsymbol{\Sigma}^P_{\Omega^c}$ in \eqref{eq:condp}.
\subsection{Can we directly use the SMG model for UQ?}
Lemma \ref{thm:smg} provides a closed-form posterior distribution for the low-rank matrix $\bm{X}$ \textit{after} observing the noisy observations $\bm{Y}_{\Omega}$. It points to a potential way for computing confidence intervals on each entry in $\bm{X}$, assuming the underlying row and column subspaces $\mathcal{U}$ and $\mathcal{V}$ are known. Of course, in practice, such subspaces are never known with certainty. One solution might be to plug in point estimates of $\mathcal{U}$ and $\mathcal{V}$ (estimated from data) within the predictive equations in Lemma \ref{thm:smg}, to directly estimate unobserved entries and their uncertainties. We investigate the efficacy of this plug-in approach via a simple numerical example.

The simulation set-up is as follows. Let $m = m_1 = m_2 = 8$ be the row and column dimensions of the matrix, and let $R=2$ be its rank. We first simulate two random orthonormal matrices $\bm{U}$ and $\bm{V}$ of size $m \times R$, via a truncated SVD on an $m \times m$ matrix with \textit{i.i.d.} $\mathcal{U}[0,1]$ entries. With $\mathcal{P_U}=\bm{UU}^T$ and $\mathcal{P_V}=\bm{VV}^T$, the ``true'' low-rank matrix is then simulated from the SMG model $\bm{X}\sim\mathcal{SMG}(\mathcal{P_U},\mathcal{P_V},\sigma^2=1,R=2)$. Finally, noisy observations are sampled via \eqref{eq:obs} with noise variance $\eta^2=0.5^2$. In total, 36 entries are observed (56.25\% of total entries), with such entries chosen uniformly at random. From this, we can obtain point estimates of the subspaces $\mathcal{U}$ and $\mathcal{V}$, by first estimating $\bm{X}$ via nuclear norm minimization (\citealp{CP2010}), a popular method for matrix completion, and then taking the row and column subspaces for this matrix estimate via SVD. These subspace estimates are then plugged into the expressions in Lemma \ref{thm:smg} for UQ. This process is then replicated for 50 times.


\begin{figure}[!t]

\begin{center} 
\subfigure[Coverage ratio using the CIs constructed via Lemma \ref{thm:smg} with plug-in subspace estimates.]{
\includegraphics[width = .49\textwidth]{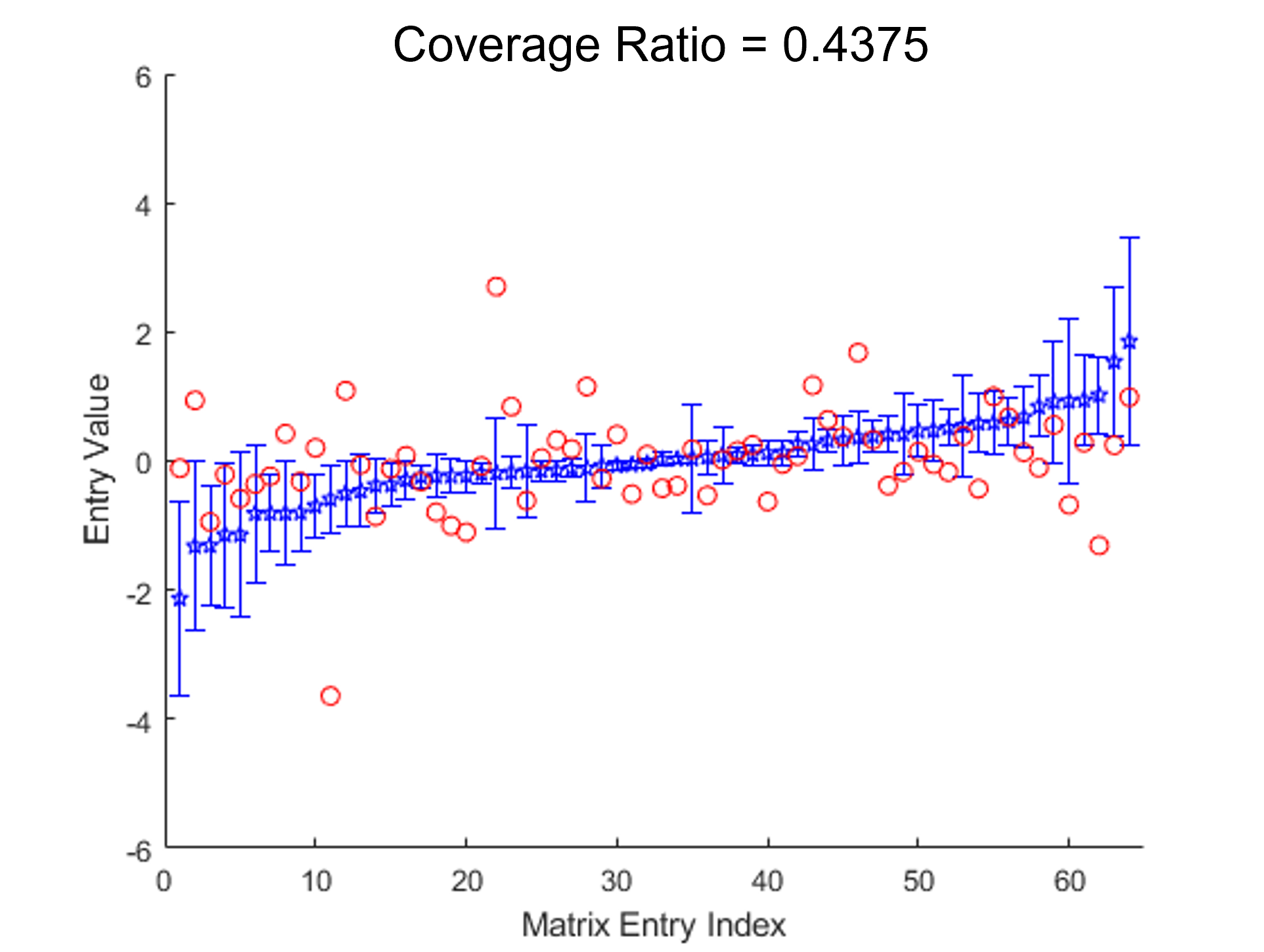}}
\hfill
\subfigure[Coverage ratio using the posterior predictive intervals from the proposed BayeSMG model..]
{\includegraphics[width = .49\textwidth]{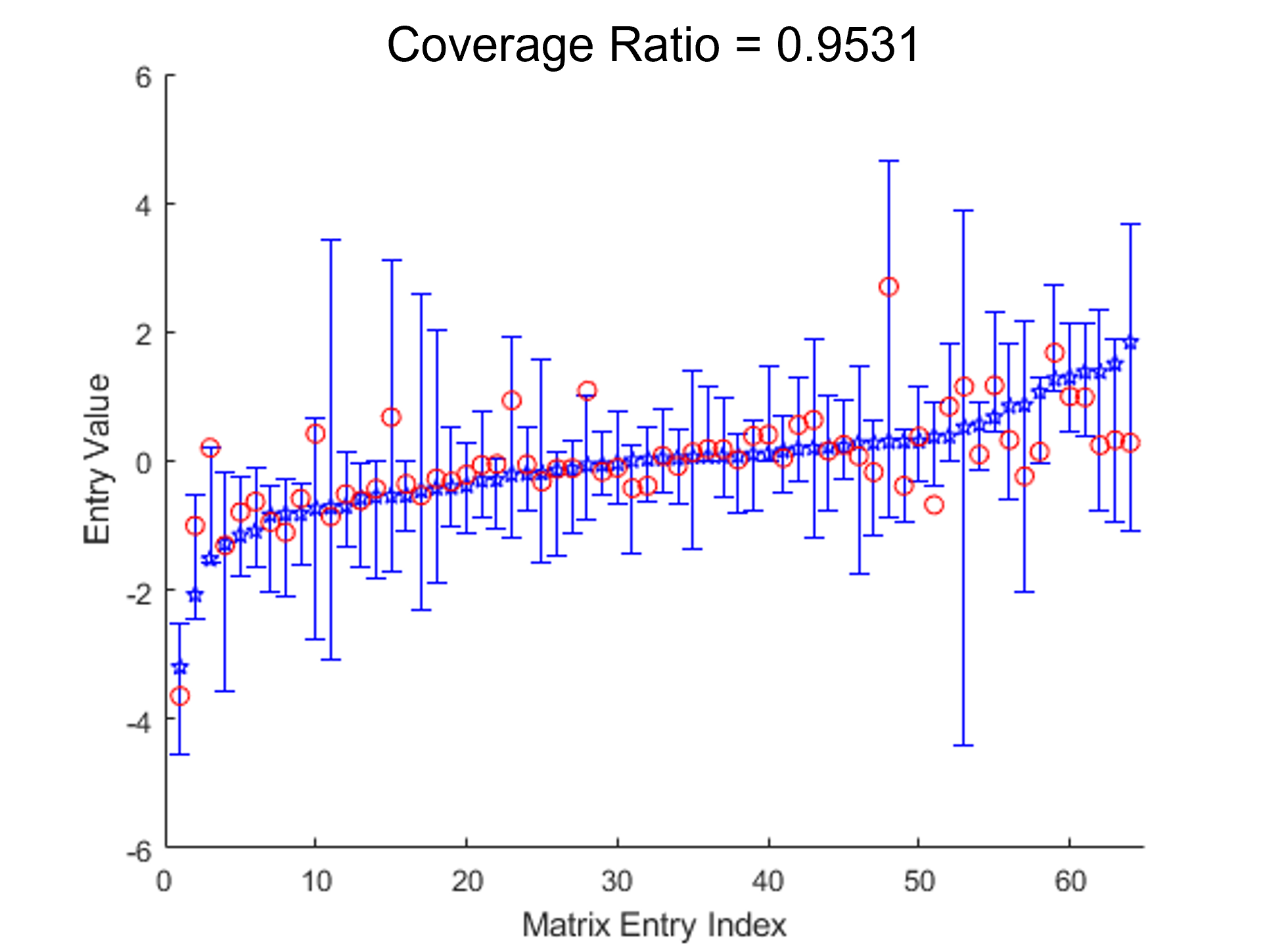}}
\vspace{-0.15in}
\caption{Plotted are the point estimates (blue points) and 95\% Confidence Intervals (blue intervals) for each matrix entry (64 in total), ordered by increasing point estimates. Red points mark the true matrix values.}
\label{fig:toy}
\vspace{-0.25in}
\end{center}
\end{figure}

Figure \ref{fig:toy}(a) plots, for a representative simulation run, the point estimates and 95\% plug-in confidence intervals (CIs) for each matrix entry using Lemma \ref{thm:smg}, with its corresponding true value marked in red. We see that these intervals provide poor coverage performance since many of the true matrix entries are not within these intervals. For this replication, the coverage ratio is only 43.8\%, and across the 50 replications, the average coverage ratio is only 46.1\%, meaning only around half of the confidence intervals cover the true entries. This poor coverage suggests that this CI approach (with plug-in subspace estimates) can significantly underestimate the underlying uncertainty of point estimates, which is unsurprising since uncertainty for subspace estimation is not incorporated when using Lemma \ref{thm:smg}. Figure \ref{fig:toy}(b) plots, for a representative simulation, the point estimates and 95\% posterior predictive intervals using the proposed BayeSMG method, which accounts for subspace uncertainty by assigning hierarchical priors on subspaces $\mathcal{U}$ and $\mathcal{V}$ from the SMG model. We see that our approach yields much better coverage: the 95\% intervals, which are now slightly wider, cover the true matrix entries well. For this replication, the coverage ratio is at 95.3\%, and across the 50 replications, the average coverage ratio is 93.9\%, which is much closer to the nominal coverage rate of 95\% than the earlier plug-in approach. This shows the proposed method can indeed provide better uncertainty quantification of $\bm{X}$ via a fully-Bayesian model specification on matrix subspaces.


\section{The BayeSMG model}\label{sec:prior}

\subsection{Model specification}
We now present the hierarchical specification for the proposed Bayesian SMG model, or BayeSMG for short. We begin by first introducing the matrix von Mises-Fisher (vMF) distribution, which will serve as prior models for the row and column orthonormal frames $\bm{U}$ and $\bm{V}$. We then present a Gibbs sampling algorithm that makes use of a reparameterization of the SMG model for efficient posterior sampling.

The matrix von Mises-Fisher distribution (\citealp{khatri1977mises,mardia2009directional}) provides a useful class of distributions on the row and column frames, which lie on a so-called Stiefel manifold. A Stiefel manifold \citep{Chi2012} consists of all orthonormal subspaces of rank $R$ in the space of $\mathbb{R}^m$; this is denoted as $\mathcal{V}_{R,m}$ hereafter. 
The matrix vMF distribution assumes the following probability density function of matrix $\bm{W}$ on $\mathcal{V}_{R,m}$:
\begin{equation}
p(\bm{W};m,R,\bm{F}) = \left[ {}_0 F_1 \left(; \frac{m}{2};\frac{\bm{F}^T\bm{F}}{4}\right) \right]^{-1} \; \text{etr}(\bm{F}^T \bm{W}), \quad \bm{W} \in \mathcal{V}_{R,m},
\label{eq:mf}
\end{equation}
where ${}_0 F_1 (;\cdot;\cdot)$ is the hypergeometric function, and $\bm{F} \in \mathbb{R}^{m \times R}$ is the concentration matrix. We denote this distribution by $\bm{W} \sim \mathcal{MF}(m,R,\bm{F})$. The matrix vMF distribution provides conditionally conjugate priors for a wide range of multivariate models, including for cluster analysis \citep{gopal2014mises} and factor models \citep{hoff2013bayesian}. One appeal of this class of distribution is that it can be efficiently sampled. \cite{Hof2009} proposed a rejection sampling algorithm that sequentially samples each column of the matrix $\bm{W}$. Recently, \cite{jauch2020monte} presented a general simulation framework on the Stiefel manifolds using polar expansions; using such an expansion with Hamiltonian Monte Carlo \citep{girolami2011riemann} provides a better sampling efficiency over competing MCMC methods by an order of magnitude. We will leverage this useful family of priors via the following reparametrization of the BayeSMG model.

The following proposition gives a nice reformulation of the SMG model under uniform subspace priors on $\mathcal{U}$ and $\mathcal{V}$:
\begin{proposition}[SVD of BayeSMG]
Suppose $\bm{X} \sim \mathcal{SMG}(\mathcal{P}_{\mathcal{U}},\mathcal{P}_{\mathcal{V}},\sigma^2,R)$, with independent uniform priors $\mathcal{P}_{\mathcal{U}} \sim {U}(\mathcal{G}_{R,m_1-R})$, $\mathcal{P}_{\mathcal{V}} \sim {U}(\mathcal{G}_{R,m_2-R})$, and fixed $\sigma^2$ and $R$. Let $\bm{X} = \bm{U} \bm{D} \bm{V}^T$ be the SVD of $\bm{X}$, with singular values $\textup{diag}(\bm{D}) = (d_k)_{k=1}^R$ not necessarily in decreasing order. Then:
\begin{enumerate}
\item The singular vectors $\bm{U}$ and $\bm{V}$ follow independent priors $\mathcal{MF}(m_1,R,\bm{0})$ and $\mathcal{MF}(m_2,R,\bm{0})$, respectively.
\item The singular values $\textup{diag}(\bm{D}) = (d_k)_{k=1}^R$ follow the repulsed normal distribution, with density:
\begin{equation}
\frac{1}{Z_R(2\pi\sigma^2)^{R/2}} \exp\left\{ -\frac{1}{2 \sigma^2} \sum_{k=1}^R d_k^2 \right\} \prod_{\substack{k,l=1\\k < l}}^R |d_k^2 - d_l^2|, \; d_k > 0, \, k = 1, \cdots, R.
\label{eq:ql0}
\end{equation}
\end{enumerate}
\label{prop:svd}
\end{proposition}
\noindent 
The proof of this proposition is provided in the supplementary section. The first part of the proposition shows that the use of uniform priors on the projection matrices $\mathcal{P}_{\mathcal{U}}$ and $\mathcal{P}_{\mathcal{V}}$ corresponds to independent $\mathcal{MF}(m_1,R,\bm{0})$ and $\mathcal{MF}(m_2,R,\bm{0})$ priors for the singular vectors $\bm{U}$ and $\bm{V}$, which are uniform priors on the Stiefel manifolds $\mathcal{V}_{R,m_1}$ and $\mathcal{V}_{R,m_2}$, respectively. The second part shows that the singular values in $\bm{D}$ follow the repulsed normal distribution, which is closely connected with the distribution of singular values for a Gaussian ensemble \citep{She2001}.

This proposition then motivates the following reparametrization of the BayeSMG model:
\begin{equation}
\bm{X} = \bm{U}\bm{D}\bm{V}^T, \;\; \bm{U} \sim \mathcal{MF}(m_1,R,\bm{F}_1), \;\; \bm{V} \sim \mathcal{MF}(m_2,R,\bm{F}_2), \;\; \text{diag}(\bm{D}) \sim \mathcal{RN}(\bm{0},\sigma^2),
\label{eq:reparam}
\end{equation}
where $\mathcal{RN}(\bm{0},\sigma^2)$ is the repulsed normal distribution in \eqref{eq:ql0}, and the priors on $\bm{U}$, $\bm{V}$ and $\bm{D}$ are independently specified. When little is known a priori on matrix subspaces, one can set the concentration matrices as $\bm{F}_1 = \bm{F}_2 = \bm{0}$, which provides non-informative priors on $\bm{U}$ and $\bm{V}$. In problems where some prior information is available on matrix subspaces, one can elicit a good choice of prior parameters for the vMF priors via a moment matching approach \citep{wang2009general}. We show in the next section that this reparametrization allows for a Gibbs sampling algorithm which makes use of conditionally conjugate priors for efficient posterior sampling.

Finally, we complete the Bayesian specification by assigning the following priors on the variance parameters $\sigma^2$ and $\eta^2$:
\begin{align}
\begin{split}
[\sigma^2] \sim IG(\alpha_{\sigma^2},\beta_{\sigma^2}), \quad [\eta^2] \sim IG(\alpha_{\eta^2},\beta_{\eta^2}),
\label{eq:prior}
\end{split}
\end{align}
where $IG(\alpha,\beta)$ is the Inverse-Gamma distribution with shape and rate parameters $\alpha$ and $\beta$. Table \ref{tbl:spec} summarizes the full Bayesian model specification for BayeSMG.

\begin{table}
\begin{tabular}{ l l l }
\toprule
 \multicolumn{1}{c}{\textit{Model}} &  \multicolumn{1}{c}{\textit{Distribution}}\\
\toprule
\textbf{Observations} & $[\bm{Y}_{\Omega}|\bm{X},\eta^2]$: $Y_{i,j} \distas{i.i.d.} \mathcal{N}(X_{i,j}, \eta^2)$\\
\hline
\textbf{Low-rank matrix} & $[\bm{X}|\mathcal{P}_{\mathcal{U}}, \mathcal{P}_{\mathcal{V}},\sigma^2]:\bm{X} \sim \mathcal{SMG}(\mathcal{P}_{\mathcal{U}}, \mathcal{P}_{\mathcal{V}},\sigma^2,R) $\\
(equivalently) & $[\bm{X}|\bm{U}, \bm{V},\sigma^2]:\bm{X} = \bm{U} \bm{D} \bm{V}^T$, $\text{diag}\{\bm{D}\} \sim \mathcal{RN}(\bm{0},\sigma^2)$\\
\hline
\textbf{Priors} & $[\mathcal{P}_{\mathcal{U}}, \mathcal{P}_{\mathcal{V}},\sigma^2,\eta^2] = [\mathcal{P}_{\mathcal{U}}]\;[\mathcal{P}_{\mathcal{V}}]\;[\eta^2][\sigma^2] $ \\
\quad Matrix subspaces & \quad $[\mathcal{P}_{\mathcal{U}}] \sim {U}(\mathcal{G}_{R,m_1-R})$\\
 & \quad $[\mathcal{P}_{\mathcal{V}}] \sim {U}(\mathcal{G}_{R,m_2-R})$\\
\quad Matrix variance & \quad $[\sigma^2] \sim IG(\alpha_{\sigma^2}, \beta_{\sigma^2})$\\
\quad Noise variance& \quad $[\eta^2] \sim IG(\alpha_{\eta^2},\beta_{\eta^2})$\\
\toprule
\end{tabular}
\caption{Model specification for BayeSMG.}
\normalsize
\vspace{-0.3cm}
\label{tbl:spec}
\end{table}
\subsection{Posterior sampling}
\label{sec:post}
Using the reparametrized model \eqref{eq:reparam}, we now present a subspace Gibbs sampler for posterior sampling on the BayeSMG model, specifically on the parameters $\Theta = \{\bm{U},\bm{D},\bm{V},\sigma^2\}$ given partial and noisy observations $\bm{Y}_{\Omega}$. We first introduce the sampler under \textit{complete} observation of the noisy matrix $\bm{Y}$, then describe a data imputation procedure for posterior sampling under \textit{partial} observations $\bm{Y}_{\Omega}$.

Consider first the setting where complete observations on $\bm{Y}$ are obtained. It can then be shown (see supplementary material for a full derivation) that the full conditional distributions of $\bm{U}$, $\bm{D}$, $\bm{V}$ and $\sigma^2$ take the form:
\begin{equation}
\begin{split}
[\bm{U}|\bm{D},\bm{V},\bm{Y},\sigma^2,\eta^2] & \sim \mathcal{MF}(m_1, R, \bm{Y}\bm{V}\bm{D}/\eta^2 + \bm{F}_1),\\
[\bm{V}|\bm{D},\bm{U},\bm{Y},\sigma^2,\eta^2] & \sim \mathcal{MF}(m_2, R, \bm{Y}^T\bm{U}\bm{D}/\eta^2 + \bm{F}_2),\\
[\bm{D}|\bm{U},\bm{V},\bm{Y},\sigma^2,\eta^2] & \sim \mathcal{RN}\left( \sigma^2 \text{diag}(\bm{U}^T \bm{Y} \bm{V}) / (\eta^2 + \sigma^2), \eta^2 \sigma^2 / (\eta^2 + \sigma^2) \right),\\
[\sigma^2|\bm{U},\bm{D},\bm{V},\bm{Y},\eta^2] & \sim IG(\alpha_{\sigma^2} + R/2, \beta_{\sigma^2} + \text{tr}(\bm{D}^2)/2),\\
[\eta^2|\bm{U},\bm{D},\bm{V},\bm{Y},\sigma^2] & \sim IG( \alpha_{\eta^2} + m_1 m_2 / 2, \beta_{\eta^2} + \|\bm{Y} - \bm{U}\bm{D}\bm{V}^T\|_F^2/2).
\label{eq:gibbscomplete}
\end{split}
\end{equation}
Here, $\| \bm{M} \|_F=\sqrt{\sum_{i,j} M_{i,j}^2}$ is the Frobenius norm of matrix $\bm{M}$. One can then perform the above full conditional updates cyclically for posterior sampling on $[\Theta|\bm{Y}]$ via Gibbs sampling. These full conditional sampling steps are related to the Gibbs sampler proposed in \cite{hoff2007model} for probabilistic SVD. As mentioned previously, there are efficient sampling algorithms for the matrix vMF distribution \citep{Hof2009,jauch2020monte}, which enable efficient full conditional sampling on $\bm{U}$ and $\bm{V}$. The full conditional distribution of $\bm{D}$ follows the aforementioned repulsed normal distribution with a location shift of $\boldsymbol{\mu}$ (denoted as $\mathcal{RN}(\boldsymbol{\mu},\delta^2)$), with density:
\begin{equation}
\frac{1}{Z_R(2\pi\delta^2)^{R/2}} \exp\left\{ -\frac{1}{2 \delta^2} \sum_{k=1}^R (d_k - \mu_k)^2 \right\} \prod_{\substack{k,l=1; k < l}}^R |d_k^2 - d_l^2|,
\label{eq:gql}
\vspace{-0.1in}
\end{equation}
where $d_k > 0, k = 1, \cdots R$. We have found that this can be quite efficiently sampled via a Metropolis-Hastings sampler \citep{metropolis1953equation}, with an ``independent'' proposal distribution (i.e., independent of the current state) set as a non-central, multivariate $t$-distribution with mean vector $\boldsymbol{\mu}$ and scale parameter $\delta$.


Consider now the setting where only partial noisy observations $\bm{Y}_{\Omega}$ are available. We describe a posterior sampling algorithm for $[\Theta|\bm{Y}_{\Omega}]$, which makes use of a modification on the above Gibbs sampler on $[\Theta|\bm{Y}]$. The idea is to first sample from the joint distribution $[\Theta, \bm{Y}_{\Omega^c}|\bm{Y}_{\Omega}]$ of both the parameters $\Theta$ and unobserved noisy entries $\bm{Y}_{\Omega^c}$, then take only the marginal samples of parameters $\Theta$. With an initialization of $\Theta = \Theta'$, the joint distribution $[\Theta, \bm{Y}_{\Omega^c}|\bm{Y}_{\Omega}]$ can be sampled via the following Gibbs sampling steps:
\begin{enumerate}
    \item[(i)] Draw one sample from $[\bm{Y}_{\Omega^c}| \bm{Y}_{\Omega},\Theta']$. Since the missing entries $\bm{Y}_{\Omega^c}$ is assumed to be conditionally independent of the observed entries $\bm{Y}_{\Omega}$ given $\bm{X} = \bm{U} \bm{D} \bm{V}^T$, this is equivalent to sampling $[\bm{Y}_{\Omega^c} | \bm{X}]$, which amounts to simulating the observation noise in $\bm{Y}_{\Omega^c}$ given ground truth $\bm{X}_{\Omega^c}$.
    \vspace{-0.1in}
    \item[(ii)] Draw one sample $\Theta'$ from the posterior distribution $[\Theta| \bm{Y}_{\Omega^c}, \bm{Y}_{\Omega}] = [\Theta| \bm{Y}]$ via the Gibbs sampling steps in \eqref{eq:gibbscomplete}.
\end{enumerate}
Step (i) can be viewed as a data imputation step, which imputes missing entries in the noisy matrix $\bm{Y}$. Step (ii) performs the earlier posterior sampling steps for parameters $\Theta$ given the full noisy matrix $\bm{Y}$.

It is worth noting that step (i) depends on an implicit assumption that the entries are either completely missing at random (CMAR) or missing at random (MAR); see \cite{little2019statistical} for further discussion on missing data modeling. When the entries are missing not at random (MNAR), the sampling of $[\bm{Y}_{\Omega^c}| \bm{Y}_{\Omega},\Theta']$ can become much more complicated, since it would depend on the underlying MNAR mechanism for missing entries. One approach is to adopt a probabilistic model for the MNAR entries (see, e.g., \citealp{hernandez2014probabilistic} for one such model), then sample $[\bm{Y}_{\Omega^c}| \bm{Y}_{\Omega},\Theta']$ given this model. There are, however, several limitations to this approach: (i) the conditional distribution $[\bm{Y}_{\Omega^c}| \bm{Y}_{\Omega},\Theta']$ may be computationally expensive to sample from in the MNAR setting, and (ii) in the case of misspecification for the MNAR model, the resulting recovery of the matrix $\bm{X}$ can be highly biased and inaccurate. In the absence of prior information on how matrix entries are missing (which is the case in many applications), it may be preferable to adopt Algorithm \ref{alg:gibbs} for posterior inference. We will show later (in Section \ref{sec:inpainting}) that the BayeSMG is empirically robust to mild violations of this implicit MAR assumption for missing entries.

\begin{algorithm}[!t]
\caption{\texttt{BayeSMG}$(\bm{Y}_{\Omega},R,\bm{F}_1, \bm{F}_2,\alpha_{\sigma^2},\beta_{\sigma^2},\alpha_{\eta^2},\beta_{\eta^2})$: Gibbs sampler for BayeSMG}
\label{alg:gibbs}
\vspace{0.2cm}
\textit{Initialization}:\hfill
\vspace{-0.1cm}
\bi
\item Complete $\bm{X}_{[0]}$ from $\bm{Y}_{\Omega}$ via nuclear-norm minimization in \eqref{eq:nuc}.
\item Initialize $[\bm{U}_{[0]},\bm{D}_{[0]}, \bm{V}_{[0]}] \leftarrow \text{svd}(\bm{X}_{[0]})$ and $\sigma^2_{[0]} > 0$.
\ei
\vspace{-0.1cm}
\textit{{Gibbs sampling}}:\hfill $T$ - total samples
\begin{flushleft}
\For{$t = 1, \ldots, T$ \hfill}{
\bi
\vspace{-0.3cm}
\item Set $\bm{X}_{[t]} \leftarrow \bm{U}_{[t-1]}\bm{D}_{[t-1]}\bm{V}_{[t-1]}^T$
\item Impute missing entries $\bm{Y}_{\Omega^c}$ by sampling $Y_{i,j} \distas{i.i.d.} X_{[t],i,j} + \mathcal{N}(0,\eta^2), \quad (i,j) \in \Omega^c$.
\item Sample $\bm{U}_{[t]} \sim \mathcal{MF}(m_1,R,{\bm{Y} \bm{V}_{[t-1]} \bm{D}_{[t-1]}}/{\eta^2_{[t-1]}} + \bm{F}_1)$.
\item Sample $\bm{V}_{[t]} \sim \mathcal{MF}(m_2,R,\bm{Y}^T \bm{U}_{[t]} \bm{D}_{[t-1]}/\eta^2_{[t-1]} + \bm{F}_2)$.
\item Sample $\bm{D}_{[t]} \sim \mathcal{RN}\left(\frac{\sigma^2_{[t-1]} \text{diag}(\bm{U}_{[t]}^T\bm{Y}\bm{V}_{[t]})}{(\eta^2_{[t-1]}+\sigma^2_{[t-1]})},\frac{\eta^2_{[t-1]} \sigma^2_{[t-1]}}{ (\eta^2_{[t-1]} + \sigma^2_{[t-1]})} \right)$.
\item Sample $\sigma^2_{[t]} \sim IG( \alpha_{\sigma^2} + R/ 2, \beta_{\sigma^2} + \text{tr}(\bm{D}_{[t]}^2) / 2 )$.
\item Sample $\eta^2_{[t]} \sim IG( \alpha_{\eta^2} + m_1 m_2 / 2, \beta_{\eta^2} + \|\bm{Y} - \bm{U}_{[t]}\bm{D}_{[t]}\bm{V}_{[t]}^T\|_F^2/2)$.
\ei
\vspace{-0.2cm}
}
\KwOut{Return posterior samples $\{(\bm{X}_{[t]}, \bm{U}_{[t]}, \bm{D}_{[t]}, \bm{V}_{[t]}, \sigma^2_{[t]}, \eta^2_{[t]})\}_{t=1}^T$.\hfill}
\end{flushleft}
\end{algorithm}

Algorithm \ref{alg:gibbs} summarizes the above steps for the posterior sampling algorithm. The algorithm is first initialized with estimates  $\bm{U}_{[0]}$, $\bm{D}_{[0]}$, and $\bm{V}_{[0]}$ obtained from a nuclear-norm completion of $\bm{X}$ \citep{Cea2012}, and $\sigma_{[0]}^2$ is randomly initialized from the prior \eqref{eq:prior}. Next, the missing noisy entries $\bm{Y}_{\Omega^c}$ are imputed using step (i), then a posterior draw is made using step (ii) via the Gibbs steps in \eqref{eq:gibbscomplete}. This is then iterated until a desired number of posterior samples is obtained. Using the posterior samples of $(\bm{U}_{[t]},\bm{D}_{[t]},\bm{V}_{[t]})$ at each iteration $t$, we can obtain a sample $\bm{X}_{[t]}=\bm{U}_{[t]}\bm{D}_{[t]}\bm{V}_{[t]}^T$ from the desired posterior distribution $[\bm{X}|\bm{Y}_{\Omega}]$. These posterior samples $\{\bm{X}_{[t]}\}_{t=1}^T$ can then be used for the target goal of uncertainty quantification: the mean of such samples provides a point estimate $\hat{\bm{X}}$ for the recovered matrix, and its variability around $\hat{\bm{X}}$ provides a measure of uncertainty for this recovery.

While the computational complexity of this algorithm is difficult to establish given the complex manifold sampling steps, we found this posterior sampler to be quite efficient and scalable in practice. For a relatively large $256 \times 256$ matrix, the sampler takes around 1 minute to generate $T=1000$ samples on a standard laptop computer (Intel i7 CPU and 16GB RAM), which is quite efficient given the size of the matrix. We will report computation times for larger matrices in the numerical studies later.

\subsection{Inference on matrix rank}
\label{sec:rank}

The BayeSMG model as presented above assumes the rank of the matrix $\bm{X}$ is known, which is often not the case in practice. There has been some literature on this problem of rank estimation for matrix inference. \cite{shapiro2018matrix} investigates a lower bound of the matrix rank needed for the matrix completion problem to be stable. \cite{hoff2007model} proposes a Bayesian dimension selection method that models the dimension of matrix subspaces via a singular value decomposition (SVD), thus allowing for a Gibbs sampler for sampling the matrix singular vectors, singular values, and rank. While one can conceptually adopt a similar fully Bayesian approach for rank $R$ here, we have found such an approach to be too computationally expensive for the high-dimensional matrices in later numerical experiments, where $m_1$ and $m_2$ can be on the order of thousands. This is because Algorithm \ref{alg:gibbs} needs to be performed for each choice of rank $R$, which can be expensive for large $m_1$ and $m_2$. For such high-dimensional applications, we instead favor the following maximum a posteriori (MAP) approach for rank inference, which sacrifices a richer quantification of uncertainty for computational efficiency and scalability.


Consider the MAP estimate of the unknown matrix $\bm{X}$, which can formulated as:
\begin{equation}
\tilde{\bm{X}} = \argmax_{\bm{X} \in \mathbb{R}^{m_1 \times m_2}} [\bm{Y}_{\Omega}|\bm{X}][\bm{X}|R][R].
\label{eq:map}
\end{equation}
Here, $[\bm{X}|R]$ follows the BayeSMG prior specification \eqref{eq:reparam} given matrix rank $R$, and $[R]$ is a prior distribution assigned on matrix rank. Under uniform subspace priors and a flat prior on $R$ over $\{1, \cdots, m_1 \wedge m_2\}$, it can be shown (see Section \ref{sec:exist} for a full derivation) that the MAP $\tilde{\bm{X}}$ can be well-approximated by the nuclear-norm formulation:
\begin{equation}
\underset{\bm{X} \in \mathbb{R}^{m_1 \times m_2}}{\textup{argmin}} \; \left( \sum_{(i,j) \in \Omega} (Y_{i,j} - X_{i,j})^2 + \lambda \|\bm{X}\|_* \right).
\label{eq:nuc}
\end{equation}
Here, $\|\bm{X}\|_*$ is the nuclear norm of $\bm{X}$ (the sum of its singular values, see \citealp{CT2010}), and $\lambda$ is a regularization parameter. The optimization problem \eqref{eq:nuc} can be efficiently solved via convex optimization algorithms (see Section \ref{sec:lit} for further details).

In practice, $\lambda$ can be estimated via cross-validation \citep{friedman2017elements} on the observed entries $\bm{Y}_\Omega$. We first divide these entries into multiple folds. For each fold, we first use nuclear-norm minimization \eqref{eq:nuc} to estimate the entries of the particular fold. Then we compute the cross-validation error for these estimates. We then select the optimal tuning parameter $\lambda^*$ such that it is the $\lambda$ that minimizes the sum of these cross-validation errors for all folds.

With this estimate $\lambda^*$, an (approximate) MAP estimate $\tilde{\bm{X}}$ can be obtained by solving \eqref{eq:nuc} with $\lambda = \lambda^*$. This in turn yields an approximate MAP estimate of $R$ via the rank of the matrix estimate $\tilde{\bm{X}}$. Finally, this rank estimate can be plugged into Algorithm \ref{alg:gibbs} for uncertainty quantification on matrix $\bm{X}$. For high-dimensional problems with either $m_1$ or $m_2$ large, this plug-in MAP approach for rank estimation can yield significant computational savings over a fully Bayesian treatment.

\section{Theoretical insights}
\label{sec:uq}
We now provide some theoretical insights on the BayeSMG model. We first discuss an interesting link between the maximum-a-posterior (MAP) estimator and regularized estimators in the literature, then present a connection between model uncertainty from the BayeSMG model and coherence, which is then used to prove an error monotonicity result on uncertainty quantification.
\subsection{Connection to Regularized Estimators}
\label{sec:exist}

The following lemma reveals a connection between the BayeSMG model and existing completion methods:
\begin{lemma}[MAP estimator]
Assume the BayeSMG model in Table \ref{tbl:spec}, with $\bm{F}_1 = \bm{F}_2=\bm{0}$, $\eta^2$ and $\sigma^2$ fixed, and a uniform prior on rank $R$. Conditional on $\bm{Y}_{\Omega}$, the MAP estimator 
for $\bm{X}$ becomes
\begin{equation}
\underset{\bm{X} \in \mathbb{R}^{m_1 \times m_2}}{\textup{argmin}} \; \left( \frac{\|\bm{Y}_{\Omega}-\bm{X}_{\Omega}\|_2^2}{\eta^2} + \log(2 \pi \sigma^2) \textup{rank}^2(\bm{X}) + \frac{\|\bm{X}\|_F^2}{\sigma^2} \right),
\label{eq:matest}
\end{equation}
\normalsize
where $\|\bm{X}\|_F = \sqrt{\sum_{i,j} X_{i,j}^2}$ is the Frobenius norm of $\bm{X}$.
\label{thm:mle}
\end{lemma}

The MAP estimator $\tilde{\bm{X}}$ in \eqref{eq:matest} connects the proposed model with existing deterministic matrix completion methods (see \citealp{DR2016} and references therein). Consider the following approximation to the MAP formulation \eqref{eq:matest}. Treating $\log(2 \pi \sigma^2) \textup{rank}^2(\bm{X})$ as a Lagrange multiplier, one can view this as a constraint on $\textup{rank}^2(\bm{X})$, or equivalently, on $\textup{rank}(\bm{X})$. Replacing $\text{rank}(\bm{X})$ by its nuclear norm $\|\bm{X}\|_*$ (its tightest convex relaxation, see \citealp{Rea2010}), and treating this new constraint as a Lagrange multiplier, the optimization in \eqref{eq:matest} becomes:
\begin{equation}
\underset{\bm{X} \in \mathbb{R}^{m_1 \times m_2}}{\textup{argmin}} \;  \|\bm{Y}_{\Omega}-\bm{X}_{\Omega}\|_2^2 + \lambda \left\{ \alpha \|\bm{X}\|_* + (1-\alpha) \|\bm{X}\|_F^2 \right\},
\label{eq:el}
\end{equation}
\normalsize
for some choice of $\lambda > 0$ and $\alpha \in (0,1)$. Using \eqref{eq:el} to approximate \eqref{eq:matest}, we can then view the MAP estimator as an analogue of the \textit{elastic net} estimator \citep{ZH2005} from linear regression for noisy matrix completion.

To see the connection between the MAP estimator $\tilde{\bm{X}}$ and existing matrix completion methods, set $\alpha = 1$ in \eqref{eq:el}. The problem then reduces to the nuclear-norm formulation in \eqref{eq:nuc}, which is widely used for deterministic matrix completion \citep{CR2012,CT2010,Rec2011}. This provides an intuitive connection between the proposed Bayesian model and existing completion methods, which we leveraged earlier for efficient inference on matrix rank.

\subsection{Uncertainty and coherence}
\begin{figure}[!t]
\vspace{-0.1in}
\centering
\includegraphics[width=0.5\textwidth]{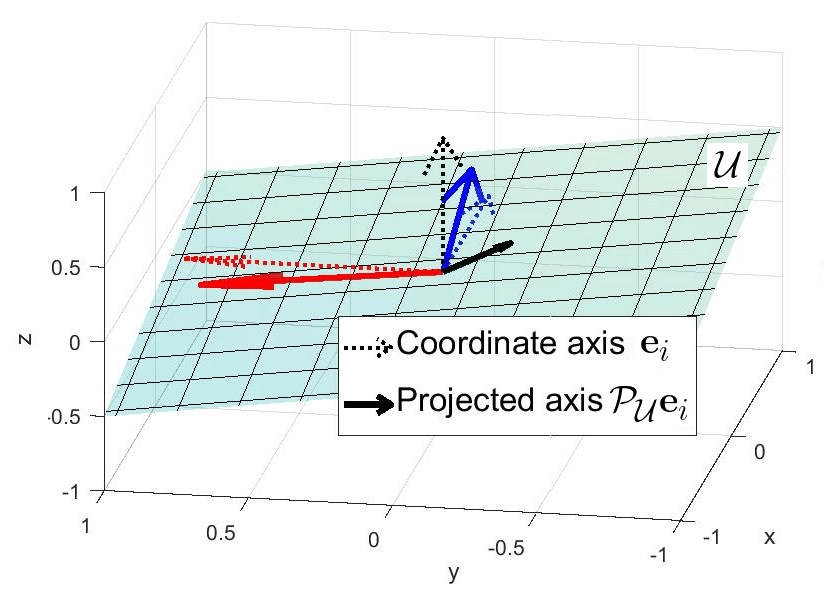}
\vspace{-0.2in}
\caption{A visualization of near-maximal coherence (red basis vector) and minimal coherence (black basis vector) for subspace $\mathcal{U}$.}
\label{fig:cohvis}
\vspace{-0.2in}
\end{figure}

Consider next the following definition of subspace coherence from \cite{CR2012}, ignoring scaling factors:
\begin{definition}[Coherence, Definition 1.2 of \citealp{CR2012}]
Let $\mathcal{U} \in \mathcal{G}_{R,m-R}$ be an $R$-plane in $\mathbb{R}^m$, and let $\mathcal{P}_{\mathcal{U}}$ be the orthogonal projection onto $\mathcal{U}$. The \textup{coherence} of subspace $\mathcal{U}$ with respect to the $i$-th basis vector, $\bm{e}_i$, is defined as $\mu_i(\mathcal{U}) := \|\mathcal{P}_{\mathcal{U}} \bm{e}_i\|_2^2$,
and the \textup{coherence} of $\mathcal{U}$ is defined as $\mu(\mathcal{U}) = \max\limits_{i = 1, \ldots, m} \mu_i(\mathcal{U})$.
\label{def:coh}
\end{definition}
\noindent In words, coherence measures how \textit{correlated} a subspace $\mathcal{U}$ is with the basis vectors $\{\bm{e}_i\}_{i=1}^m$. A large $\mu_i(\mathcal{U})$ suggests that $\mathcal{U}$ is highly correlated with the $i$-th basis vector $\bm{e}_i$, in that the projection of $\bm{e}_i$ onto $\mathcal{U}$ preserves much of its original length; a small value of $\mu_i(\mathcal{U})$ suggests that $\mathcal{U}$ is nearly orthogonal with $\bm{e}_i$, so a projection of $\bm{e}_i$ onto $\mathcal{U}$ loses most of its length. Figure \ref{fig:cohvis} visualizes these two cases using the projection of three basis vectors on a two-dimensional subspace $\mathcal{U}$. Note that the projection of the red vector onto $\mathcal{U}$ retains nearly unit length, so $\mathcal{U}$ has near-maximal coherence for this basis. The projection of the black vector onto $\mathcal{U}$ results in a considerable length reduction, so $\mathcal{U}$ has near-minimal coherence for this basis. The overall coherence of $\mathcal{U}$, $\mu(\mathcal{U})$, is largely due to the high coherence of the red basis vector.

In matrix completion literature, coherence is widely used to quantify the \textit{recoverability} of a low-rank matrix $\bm{X}$. Here, the same notion of coherence arises in a different context within the proposed model's uncertainty quantification. Lemma \ref{thm:smg} provides the basis for this connection. Consider first the case where no matrix entries have been observed. From Lemma \ref{thm:smg}(a), $\text{vec}(\bm{X})$ follows the degenerate Gaussian distribution $\mathcal{N}\{\bm{0},\sigma^2(\mathcal{P}_{\mathcal{V}} \otimes \mathcal{P}_{\mathcal{U}}) \}$. The variance of the $(i,j)$-th entry in $\bm{X}$ can then be shown to be:
\begin{equation}
\text{Var}(X_{i,j}) = \sigma^2 ( \bm{e}_i^T \mathcal{P}_{\mathcal{U}} \bm{e}_i )( \bm{e}_j^T \mathcal{P}_{\mathcal{V}} \bm{e}_j) = \sigma^2 \mu_i(\mathcal{U}) \mu_j(\mathcal{V}).
\label{eq:varxij}
\end{equation}
Hence, before observing data, the model uncertainty for entry $X_{i,j}$ is proportional to the product of coherences for the row and column spaces $\mathcal{U}$ and $\mathcal{V}$, corresponding to the $i$-th and the $j$-th basis vectors. Put another way, BayeSMG assigns \textit{greater variation} to matrix entries with \textit{higher} subspace coherence in either its row or column index. It is quite appealing given the original role of coherence in matrix completion, where larger row (or column) coherences imply greater ``spikiness'' for entries; our framework accounts for this by assigning greater model uncertainty to such entries.

Consider next the case where noisy entries $\bm{Y}_{\Omega}$ have been observed. Let us adopt a slightly generalized notion of coherence:
\begin{definition}[Cross-coherence]
The \textup{cross-coherence} of subspace $\mathcal{U}$ with respect to the basis vectors $\bm{e}_i$ and $\bm{e}_{i'}$ is defined as $\nu_{i,i'}(\mathcal{U}) = \bm{e}_{i'}^T \mathcal{P}_{\mathcal{U}}\bm{e}_{i}$.  \label{def:ccoh}
\end{definition}
\noindent The cross-coherence $\nu_{i,i'}(\mathcal{U})$ quantifies how correlated the basis vectors $\bm{e}_i$ and $\bm{e}_{i'}$ are, \textit{after} a projection onto $\mathcal{U}$. For example, in Figure \ref{fig:cohvis}, the pair of red / blue projected basis vectors have negative cross-coherence for $\mathcal{U}$, whereas the pair of blue / black projected vectors have positive cross-coherence. When $i = i'$, this cross-coherence reduces to the original coherence in Definition \ref{def:coh}.

Define now the cross-coherence vector $\boldsymbol{\nu}_i(\mathcal{U}) = [\nu_{i,i_n}(\mathcal{U})]_{n=1}^N \in \mathbb{R}^N$, where again $\Omega = \{(i_n,j_n)\}_{n=1}^N$. From equation \eqref{eq:condp} in Lemma \ref{thm:smg}, the conditional variance of entry $X_{i,j}$ for an unobserved index $(i,j) \in \Omega^c$ becomes:
\begin{equation}
\text{Var}(X_{i,j}|\bm{Y}_{\Omega}) = \sigma^2  \mu_i(\mathcal{U}) \mu_j(\mathcal{V})- \sigma^2  \boldsymbol{\nu}_{i,j}^T \left[\bm{R}_N(\Omega) + \gamma^2 \bm{I}\right]^{-1} \boldsymbol{\nu}_{i,j},
\label{eq:varcxij}
\end{equation}
\normalsize
where $\boldsymbol{\nu}_{i,j} := \boldsymbol{\nu}_i(\mathcal{U}) \circ \boldsymbol{\nu}_j(\mathcal{V})$, and $\circ$ denotes the entry-wise (Hadamard) product. The expression in \eqref{eq:varcxij} yields a nice interpretation. From a UQ perspective, the first term in \eqref{eq:varcxij}, $\mu_i(\mathcal{U}) \mu_j(\mathcal{V})$, is simply the unconditional uncertainty for entry $X_{i,j}$, \textit{prior} to observing data. The second term, $\boldsymbol{\nu}_{i,j}^T [\bm{R}_N(\Omega) + \gamma^2 \bm{I}]^{-1} \boldsymbol{\nu}_{i,j}$, can be viewed as the \textit{reduction} in uncertainty, \textit{after} observing the noisy entries $\bm{Y}_{\Omega}$. This uncertainty reduction is made possible by the correlation structure imposed on $\bm{X}$, via the SMG model; \eqref{eq:varcxij} also yields valuable insight in terms of subspace correlation. The first term $\mu_i(\mathcal{U}) \mu_j(\mathcal{V})$ can be seen as the joint correlation between (i) row space $\mathcal{U}$ to row index $i$, and (ii) column space $\mathcal{V}$ to column index $j$, \textit{prior} to any observations. The second term can be viewed as the portion of this correlation \textit{explained} by observed indices $\Omega$.

\subsection{Error monotonicity}
This link between coherence and uncertainty then sheds insight on expected error decay. This is based on the following proposition:

\begin{proposition}[Variance reduction]\label{thm_var_reduct}
Suppose $\bm{X}$ follows the BayeSMG model in Table \ref{tbl:spec}, with $\bm{F}_1 = \bm{F}_2 =\bm{0}$ and fixed $\sigma^2$ and $\eta^2$. Let $\bm{Y}_{\Omega}$ contain the noisy entries at $\Omega \subseteq [m_1] \times [m_2]$, and let $\bm{Y}_{\Omega \cup (i,j)}$ contain an additional noisy observation at $(i,j) \in \Omega^c$. For any index $(k,l) \in [m_1] \times [m_2]$, the expected variance of $X_{k,l}$ can be decomposed as
\begin{equation}
\mathbb{E}_{\mathcal{U},\mathcal{V}} [\textup{Var}(X_{k,l}|\bm{Y}_{\Omega \cup (i,j)})] = \mathbb{E}_{\mathcal{U},\mathcal{V}}[\textup{Var}(X_{k,l}|\bm{Y}_{\Omega})] - \mathbb{E}_{\mathcal{U},\mathcal{V}} \Big[\frac{\textup{Cov}^2(X_{k,l},X_{i,j}|\bm{Y}_{\Omega}) }{\textup{Var}(X_{i,j}|\bm{Y}_{\Omega}) + \eta^2}\Big],
\label{eq:errred}
\normalsize
\end{equation}
where $\textup{Var}(X_{k,l}|\bm{Y}_{\Omega \cup (i,j)})$ is provided in \eqref{eq:varcxij}, and
\[\textup{Cov}(X_{i,j},X_{k,l}|\bm{Y}_{\Omega}) = \sigma^2 \{\nu_{i,k}(\mathcal{U}) \nu_{j,l}(\mathcal{V}) - \boldsymbol{\nu}_{i,j}^T \left[ \bm{R}_N(\Omega) + \gamma^2 \bm{I}\right]^{-1} \boldsymbol{\nu}_{k,l}\}.\]

\label{thm:errred}
\normalsize
\end{proposition}
\noindent {\it Remark:} Proposition \ref{thm_var_reduct} shows, given observed indices $\Omega$, the reduction in uncertainty (as measured by variance) for an unobserved entry $X_{k,l}$, after observing an additional entry at index $(i,j)$. The last term in \eqref{eq:errred} quantifies this reduction, and can be interpreted as follows. For an unobserved index $(k,l) \notin \Omega \cup (i,j)$, the amount of uncertainty reduction is related to the ``signal-to-noise'' ratio, where the signal is the conditional squared-covariance between the ``unobserved'' entry $X_{k,l}$ and the  ``to-be-observed'' entry $X_{i,j}$, and the noise is the conditional variance of the ``to-be-observed'' entry.

The insight of \textit{error monotonicity} then follows:
\begin{corollary}[Error monotonicity] 
Suppose $\bm{X}$ follows the BayeSMG model in Table \ref{tbl:spec}, with $\bm{F}_1 = \bm{F}_2 =\bm{0}$ and fixed $\sigma^2$ and $\eta^2$. Let $[(i_n,j_n)]_{n=1}^{m_1 m_2} \subseteq [m_1] \times [m_2]$ be an arbitrary sampling sequence, where $(i_n,j_n) \neq (i_{n'},j_{n'})$ for $n \neq n'$. Let $X_{k,l}^P$ be the $(k,l)$-th entry of the conditional mean in \eqref{eq:condp}. Define the error term \[\epsilon^2_N(k,l) := \mathbb{E}_{\bm{X}}\left[ \left(X_{k,l} - X_{k,l}^P \right)^2 \Big|\bm{Y}_{\Omega_{1:N}}\right], \; (k, l) \in [m_1] \times [m_2].\] Then $\epsilon_{N+1}^2(k,l) \leq \epsilon_N^2(k,l)$ for any $(k, l) \in [m_1] \times [m_2]$ and $N = 1, 2, \cdots .$
\label{cor:mono}
\end{corollary}
\noindent {\it Remark:} This corollary shows that, for any sampling sequence and any index $(k,l)$, the expected squared-error in estimating $X_{k,l}$ with the conditional mean $X_{k,l}^P$ is always monotonically decreasing as more samples are collected. This is intuitive since one expects to gain greater accuracy and precision on the unknown matrix $\bm{X}$ as more entries are observed. The fact that the proposed model quantifies this monotonicity property provides a reassuring check on our UQ approach.
\section{Numerical experiments}
\label{sec:numerical}
We now investigate the performance of the proposed BayeSMG method in numerical experiments and compare it to the BPMF method \citep{SM2008}, a popular Bayesian matrix completion method in the literature.
\subsection{Synthetic data}

\begin{figure}[!t]
\centering
\subfigure[The four plots show (from left to right) the true matrix $\bm{X}$, observations $\bm{Y}_{\Omega}$, and the posterior mean estimates from BayeSMG and BPMF.]
{\includegraphics[width = \textwidth]{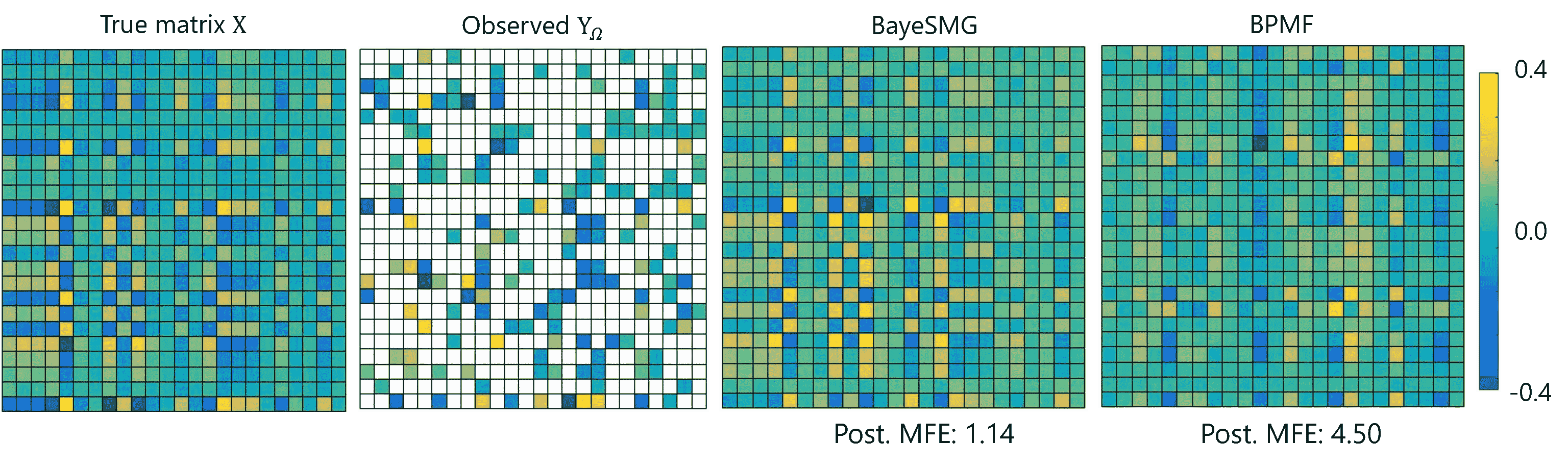}}\\
\subfigure[The left plots visualize the true row space (green) and estimated row space (blue) from BayeSMG and BPMF for the first two dimensions, with posterior MSD calculated. The right plots show the trace plots for row coherence $\mu_1$ and an unobserved entry $X_{1,1}$, for BayeSMG and BPMF, with true values dotted in red.]
{\includegraphics[width = \textwidth]{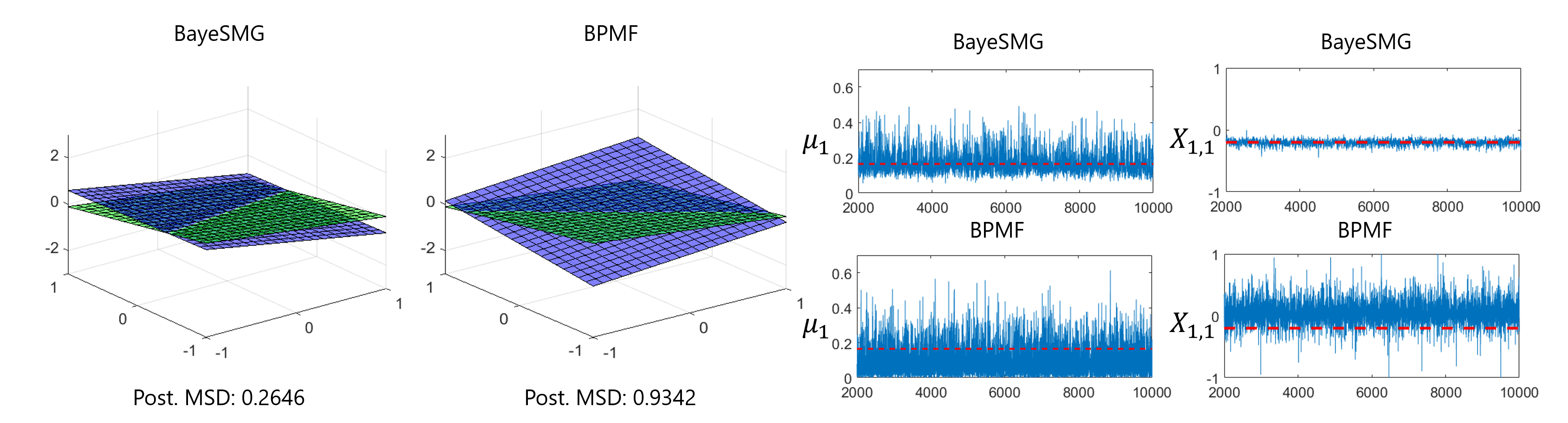}}
\vspace{-0.1in}
\caption{Recovery and UQ performance for a simulated 25$\times$25 matrix.}
\vspace{-0.2in}
\label{fig:vis}
\end{figure}

For the first numerical study, we assume the true matrix $\bm{X} \in \mathbb{R}^{24 \times 24}$ is generated from the SMG distribution, i.e., as $\bm{X} \sim \mathcal{SMG}(\mathcal{P}_{\mathcal{U}},\mathcal{P}_{\mathcal{V}},\sigma^2 = 1,R=2)$, with uniformly sampled subspaces $\mathcal{U}$ and $\mathcal{V}$. The entries are assumed to be missing-at-random and the observed entries are contaminated by noise with a variance $\eta^2=0.05^2$, which we presume to be known. The prior specifications are as follows. For BayeSMG, we assign a weakly-informative prior $\sigma^2 \sim IG(0.01,0.01)$ on the variance parameter $\sigma^2$, with non-informative manifold hyperparameters $\bm{F}_1 = \bm{F}_2 = \bm{0}$. For BPMF, we assign the recommended weak Inverse-Wishart priors on covariance matrices $\boldsymbol{\Sigma}_M\sim\text{IW}(R=2,\bm{I})$, $\boldsymbol{\Sigma}_N\sim\text{IW}(R=2,\bm{I})$. We then ran 10,000 MCMC iterations for both methods, with the first 2,000 samples taken as burn-in. Standard MCMC convergence checks were performed via trace plot inspection (see Figure \ref{fig:vis} (b)) and the Gelman-Rubin statistic \citep{gelman1992inference}.

We employ two metrics to compare the posterior contraction and UQ performance of these two methods. The first is the Mean Frobenius Error (MFE), defined as
    \[
    \text{MFE} = \frac{1}{T} \sum_{t=1}^T \|\bm{X} - \bm{X}_{[t]}\|_F.
    \]
    The MFE calculates the Frobenius norm of the difference between the posterior predictive samples $\{\bm{X}_{[t]}\}_{t=1}^T$ and the original matrix $\bm{X}$. A smaller MFE suggests better recovery and faster posterior contraction for the desired matrix $\bm{X}$. The second metric is the Mean Spectral Distance (MSD), defined as 
    \[
    \text{MSD} = \frac{1}{T} \sum_{t=1}^T d_S(\mathcal{U},\mathcal{U}_{[t]}), \quad  d_S(\mathcal{U},\mathcal{U}') := \sqrt{1 - \|\bm{U}^T\bm{U}'\|_2^2},
    \]
where $\bm{U}$ (or $\bm{U}'$) is any frame in subspace $\mathcal{U}$ (or $\mathcal{U}'$). The MSD calculates the spectral distance \citep{calderbank2015block} between the posterior samples $\{\mathcal{U}_{[t]}\}_{t=1}^T$ for the row subspaces (equivalently, $\{\mathcal{V}_{[t]}\}_{t=1}^T$ for the column subspaces) and the true row subspace $\mathcal{U}$ (equivalently, the true column subspace $\mathcal{V}$). A smaller MSD suggests better recovery and posterior contraction for matrix subspaces.

The first two plots in Figure \ref{fig:vis}(a) visualize the true matrix $\bm{X}$ and the observed $\bm{Y}_\Omega$, with 20\% of the entries observed uniformly-at-random. Here, the rank $R$ is estimated via the approximate MAP approach in Section \ref{sec:rank}. The two subsequent plots visualize the posterior mean estimates for $\bm{X}$ using BayeSMG and BPMF. We see that the BayeSMG method provides visually better recovery of the matrix $\bm{X}$, with a lower posterior MFE than the BPMF method. The first two plots in Figure \ref{fig:vis}(b) visualize the true and estimated row spaces using BayeSMG and BPMF. We again see that BayeSMG gives a visually better recovery of the row space of $\bm{X}$ (the same holds for its column space), with a lower posterior MSD than BPMF. The next two plots show the trace plots for the first-row coherence $\mu_1$ and the first matrix entry $X_{1,1}$, which is unobserved. We see that the posterior samples for BayeSMG concentrate tightly around the true coherence and matrix values, whereas the posterior samples for BPMF fluctuate much more around the truth. The above observations suggest that when the matrix is generated from the assumed prior model, BayeSMG yields much faster posterior contraction than BPMF, leading to more accurate and precise estimates of $\bm{X}$ and its subspaces. Next, we will show in the following image recovery and seismic sensor applications that the BayeSMG method provides similar improvements over BPMF via modeling and integrating subspace information.

\subsection{Image inpainting}
\label{sec:inpainting}
Image inpainting is a fundamental problem in image processing \citep{bertalmio2000image,cai2010singular}, which aims to recover and reconstruct images with missing pixels and noise corruption. It appears in numerous applications where image data are susceptible to unreliable data transmission and scratches. Take, for example, the problem of solar imaging \citep{xie2012change}. When a satellite transmits an image of the sun back to the earth, many pixels will inevitably be lost or corrupted due to the instabilities in the transmission process. The missing pixels would become a problem when the image is scaled up. In this case, the quantification of image uncertainty can be as important as the recovery, since this UQ provides insight into the quality of recovered image features in different regions. There has been some work on applying deterministic matrix completion methods for image in-painting (e.g., \citealp{xue2017depth}), but little has been done on uncertainty quantification. Our method addresses the latter goal.

We consider the aforementioned solar imaging problem, where the matrix $\bm{X}$ is a $256\times 256$ image solar flare. The pixel intensity value is encoded from 0 to 255 and represents the use of pseudo-color in the images. We then normalize pixel intensities to have zero mean and unit variance. Half of the pixels in this image are observed uniformly at random, then corrupted by Gaussian noise $\eta^2=0.05^2$. We note that, for this problem, the recovery and UQ of the row and column subspaces are of interest as well. This is because image features are often represented in the row and column spaces. Here, these subspaces may represent domain-specific, interpretable phenomena, such as different classes of solar flares, certain shapes, and sunspots. Furthermore, human eyes are typically not as sensitive to high-frequency image features; therefore, a few SVD components can often capture the vital features of an image, making its rank low. For BayeSMG and BPMF, we estimate the rank to be $R=18$ following the approximate MAP approach in Section \ref{sec:rank}, and perform 1,000 iterations of MCMC, with a burn-in period of 200. As before, MCMC convergence checks were performed via trace plot inspection and standard diagnostics.

\begin{figure}[!t]
\centering
\includegraphics[width=1\textwidth]{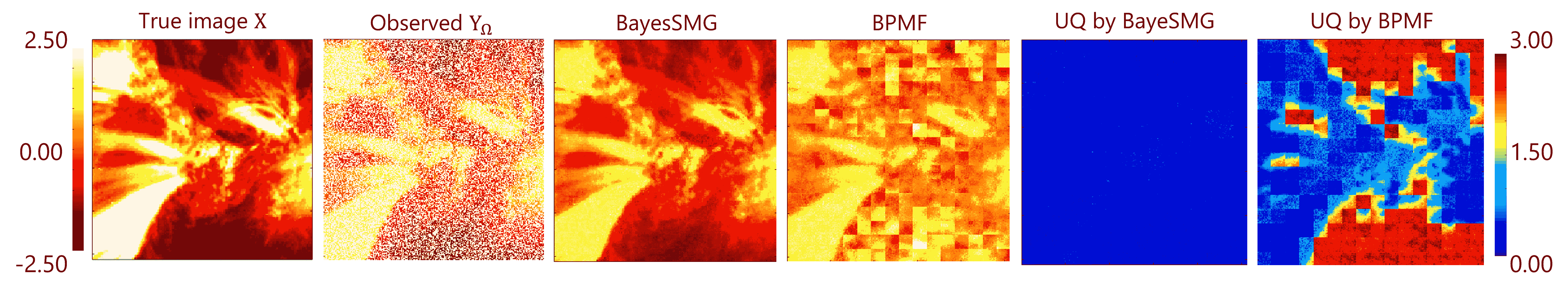}
\vspace{-0.3in}
\caption{Performance comparison between BayeSMG and BPMF on a $256\times 256$ solar flare image. The plots (from left to right) show the original image, the partially observed image with noise, the recovered images using BayeSMG and BPMF, and the widths of the entry-wise 95\% HPD intervals from BayeSMG and BPMF.}
\label{fig:flare}
\vspace{-0.2in}
\end{figure}

Figure \ref{fig:flare} shows the original solar image, its partial observations, and the recovered image using BayeSMG and BPMF via its posterior predictive mean, as well as its corresponding uncertainties via its 95\% highest posterior density (HPD) interval width \citep{hyndman1996computing}. We see that the BayeSMG method provides a much better recovery, with a noticeably lower MFE of 31.0 compared to the BPMF method (350.8). Visually, we see that the BayeSMG recovery captures the key features of the image, e.g., different types of solar flares. The BPMF recovery, on the other hand, loses much of the smaller-scale features and contains significant blocking defects. One plausible explanation is when a low-rank subspace structure is present in $\bm{X}$ (as is the case here), the proposed method can better learn and integrate this structure for improved recovery. Apart from that, an inspection of the HPD plots shows that the BayeSMG provides more accurate estimates of the recovered image, with narrow posterior HPD intervals across the whole matrix. In contrast, the BPMF is much more uncertain of its recovery: its entry-wise posterior density intervals are considerably larger, particularly for pixels with low intensities. Computation-wise, the posterior sampling for BayeSMG can be carried out within one minute on a standard laptop (Intel i7 processor with 16GB RAM), which is quite fast considering the relatively large image size.

\begin{figure}[!t]
\centering
\includegraphics[width=1\textwidth]{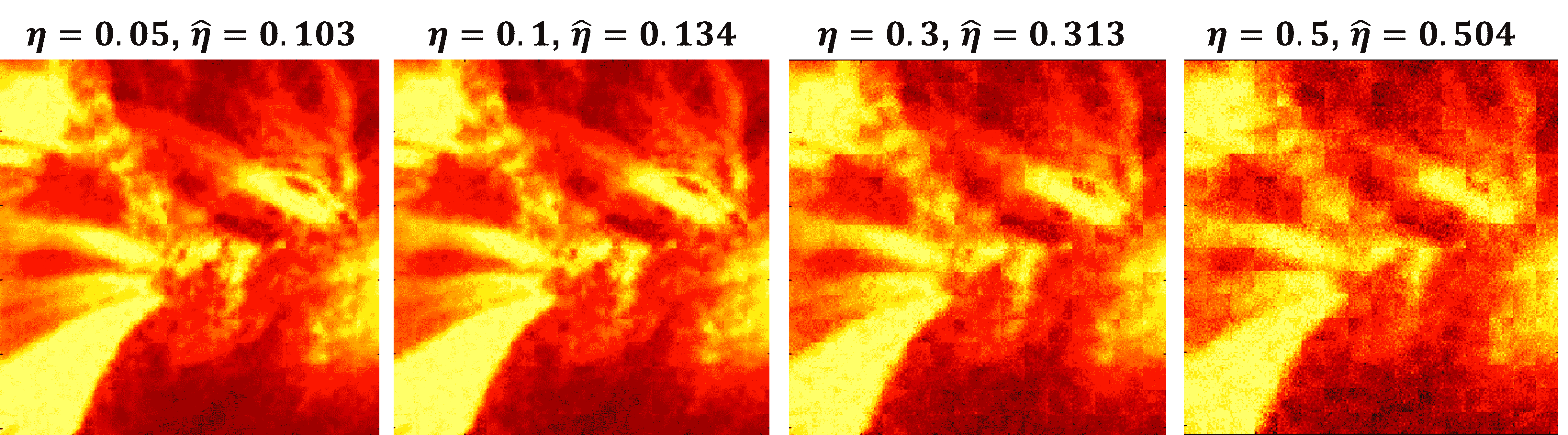}
\vspace{-0.3in}
\caption{Performance of BayeSMG on a $256\times 256$ solar flare image. The plots (from left to right) show the recovered images when the noise level $\eta=0.05,\,0.1,\,0.3,$ and $0.5$ and the estimated $\eta$ in each case by BayeSMG.}
\label{fig:noise}
\vspace{-0.2in}
\end{figure}

Additionally, we study the effect of noise on BayeSMG performance. We consider the same solar image problem, where half of the normalized matrix entries are observed and corrupted with noise. We then tested Gaussian errors with various variances $\eta^2=0.05^2,\,0.1^2,\,0.3^2$, and $0.5^2$. Figure \ref{fig:noise} shows the recovered images and the posterior estimate $\hat{\eta}$ of the noise standard deviation in each case. The MFE for the four cases are $31.00$, $35.39$, $57.48$ and $75.83$, respectively. The quality of recovery improves as noise decreases, which is as expected. For small to moderate noise levels, we see that BayeSMG yields good recovery of the solar flare image, suggesting that it is quite robust to noise. In all four cases, the posterior estimate $\hat{\eta}$ is slightly larger than the actual noise standard deviation $\eta$. One reason may be that the estimated noise level $\hat{\eta}$ captures both the true error, as well as small variations in estimating the low-rank matrix $\bm{X}$ from few observed entries. This difference becomes smaller as $\eta$ increases, which is unsurprising since the error variance would dominate the underlying low-rank matrix signal.

To demonstrate the scalability of BayeSMG, we consider next a much higher-dimensional image of the Georgia Tech campus. This image is converted to a gray-scale matrix of size $1911\times 3000$ and standardized to zero mean and unit variance. As before, half of the pixels are observed uniformly at random, then corrupted by a Gaussian noise $\eta^2=0.05^2$. To reduce computation time for posterior sampling, we fix the rank as $R = 30$ for both BayeSMG and BPMF, instead of estimating the rank using the procedure in Section \ref{sec:rank}. We run the MCMC sampler for 500 iterations after a burn-in period of 100.

\begin{figure}[!t]
\centering
\includegraphics[width=1\textwidth]{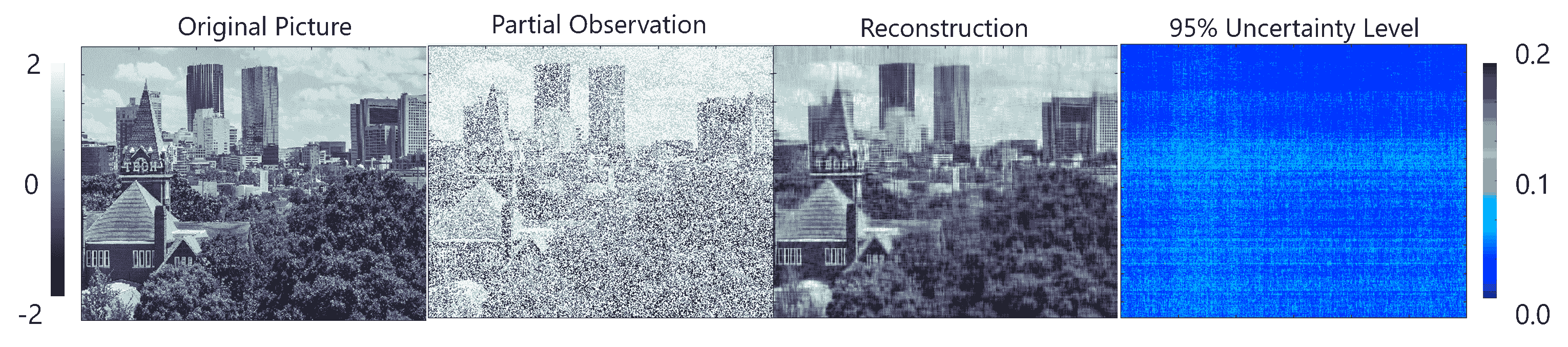}
\vspace{-0.3in}
\caption{Performance of BayeSMG on recovering a large $1911 \times 3000$ image of the Georgia Tech campus. The four plots show (from left to right) the original image, the partial observations, the recovered image using BayeSMG, and the widths of the entry-wise 95\% HPD intervals from BayeSMG.}
\label{fig:GT}
\vspace{-0.2in}
\end{figure}
Figure \ref{fig:GT} shows the true image, its partial observations, and the recovered image from BayeSMG as well as its corresponding uncertainty. The MFE of this recovery is 1005.1, which is again noticeably smaller than that for the BPMF recovery (3004.8). We see that the recovered BayeSMG image captures the original image's main features, which shows that the proposed method can learn and integrate the subspace structure for recovery. 
As before, the BayeSMG is quite confident of this completion, with narrow posterior HPD intervals over all pixels. 
Despite this being a much larger image, we can still carry out BayeSMG on the same standard laptop, albeit with a time of close to two hours. It suggests that the proposed method can yield effective probabilistic matrix recovery in high-dimensional settings.

Recall from Section \ref{sec:post} that the proposed posterior sampler for BayeSMG implicitly assumes the matrix entries are missing at random. To see how robust BayeSMG is to slight deviations from this MAR assumption, we investigate the recovery performance of BayeSMG for a $256\times 256$ lighthouse image, where the entries are missing in a not-at-random setting. In particular, we consider the MNAR case where image pixels with a higher intensity value (i.e., darker) are more likely to be observed, and pixels with a lower intensity value (i.e., lighter) are more likely to be missing. Here, 40\% of the entries with intensities higher than the population median are observed randomly, 25\% of entries with intensities equal to the median are observed randomly, and 10\% of remaining entries are observed randomly. Overall, around 25.1\% of image pixels are observed using this scheme, but the probability of missing for a single pixel depends on the true pixel intensity.

\begin{figure}[!t]
\centering
\includegraphics[width=1\textwidth]{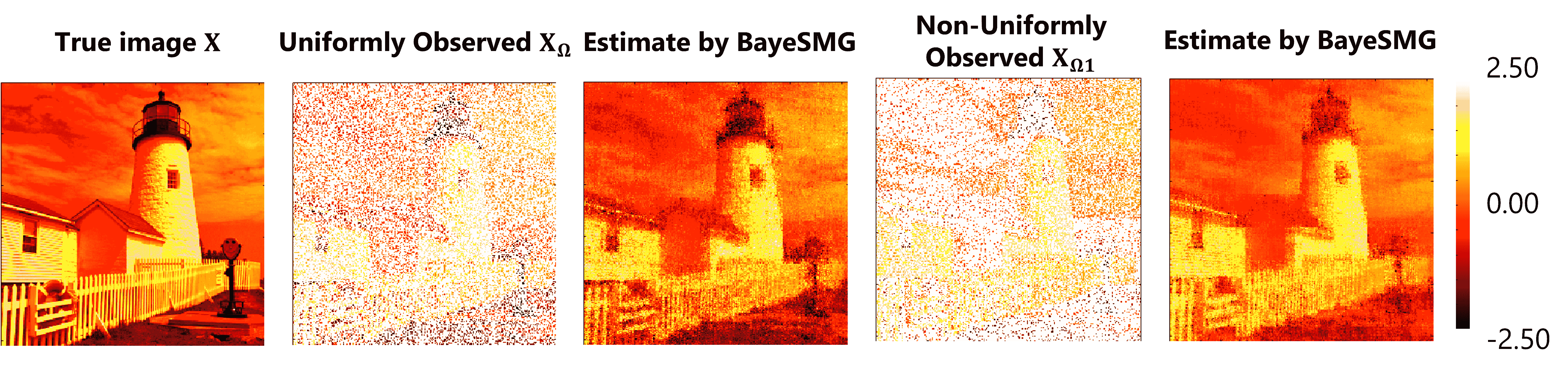}
\vspace{-0.2in}
\caption{Performance of BayeSMG on MNAR image pixels. In the first row, the first image is the original matrix, the second is the noisy matrix with entries sampled uniformly at random (MAR), and the third is its recovery estimate via the posterior mean of BayeSMG. In the second row, the first image is the noisy matrix with entries sampled MNAR, and the second image is its recovery estimate via BayeSMG.}
\label{fig:mixed-up}
\vspace{-0.2in}
\end{figure}

Figure \ref{fig:mixed-up} shows the sampled image pixels for this MNAR setting with its corresponding image recovery via the posterior mean of the BayeSMG method. For comparison, we also show the sampled pixels under an MCAR setting (where every entry is observed independently with probability 25\%), with its corresponding image recovery via BayeSMG. We estimate the ranks in both scenarios via the procedure in Section \ref{sec:rank}. For the MNAR case, the MFE is $154.35$, compared to an MFE of $148.33$ for the MCAR case. While the error is slightly higher for the MNAR case (around 4\% larger), we see from Figure \ref{fig:mixed-up} that there is little discernible difference visually between the recovered images in both cases. It suggests that the proposed BayeSMG sampler appears to be quite robust to mild violations of the implicit missing-at-random assumption for Algorithm \ref{alg:gibbs}. However, if prior information on the MNAR nature of the missing entries is known, then we can integrate such information within BayeSMG, yielding further improvements in recovery performance (see Section \ref{sec:post}).

\section{Seismic sensor network recovery}
\label{sec:seismic}
\begin{figure}[!t]
\begin{center} 
\begin{tabular}{cc}
\includegraphics[width = .43\textwidth]{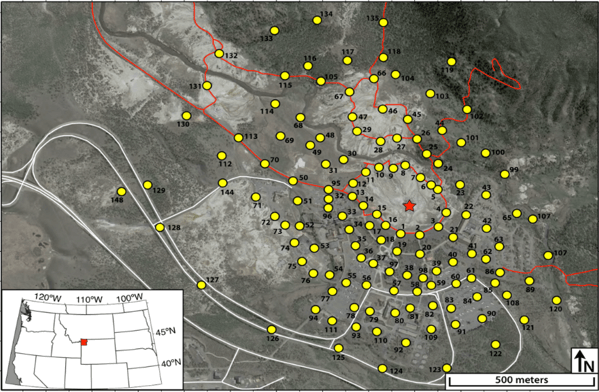}
&
\includegraphics[width = .4\textwidth]{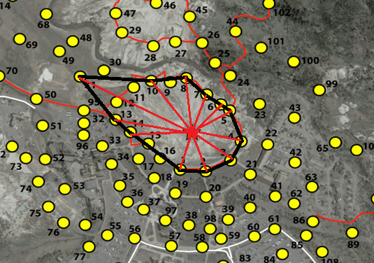}\\
(a) & (b)
\end{tabular}
\caption{The location of all 133 sensors near the geyser in Yellowstone National Park. The yellow circles indicate the sensors and the red pentagram indicates the location of the geyser. (a) shows the distribution of all 133 sensors over the region close to the geyser (see \citealp{wu2017anatomy} for details); (b) shows the locations of the 12 most significant sensors and their relative direction from each other.}
\label{fig:geyser}
\vspace{-0.2in}
\end{center}
\end{figure}

Seismic imaging is applied widely for finding oil and natural gas beneath the surface of the earth.  Ambient Noise Seismic Imaging \citep{bensen2007processing} is a relatively new technique for seismic imaging with great potential. It uses ``ambient noises'' instead of actively collected signals and is non-invasive to the environment (compared to the traditional active imaging techniques). ANSI has proved useful for imaging shallow earth structures; it utilizes pairwise cross-correlation function between signals recorded by seismic sensors followed by time-frequency analysis. From these cross-correlations, we can determine the time delay between each pair of sensors. These pairwise time delays are then combined into a data matrix, which is useful for further seismic studies. In a recent study \citep{xu2019low} on the Old Faithful geyser at Yellowstone National Park, 133 sensors were deployed in its vicinity to collect ambient noise signals for investigating geological structures. Figure \ref{fig:geyser}(a) shows the locations of these sensors. 

\begin{figure}[!t]
\centering
\includegraphics[width=1\textwidth]{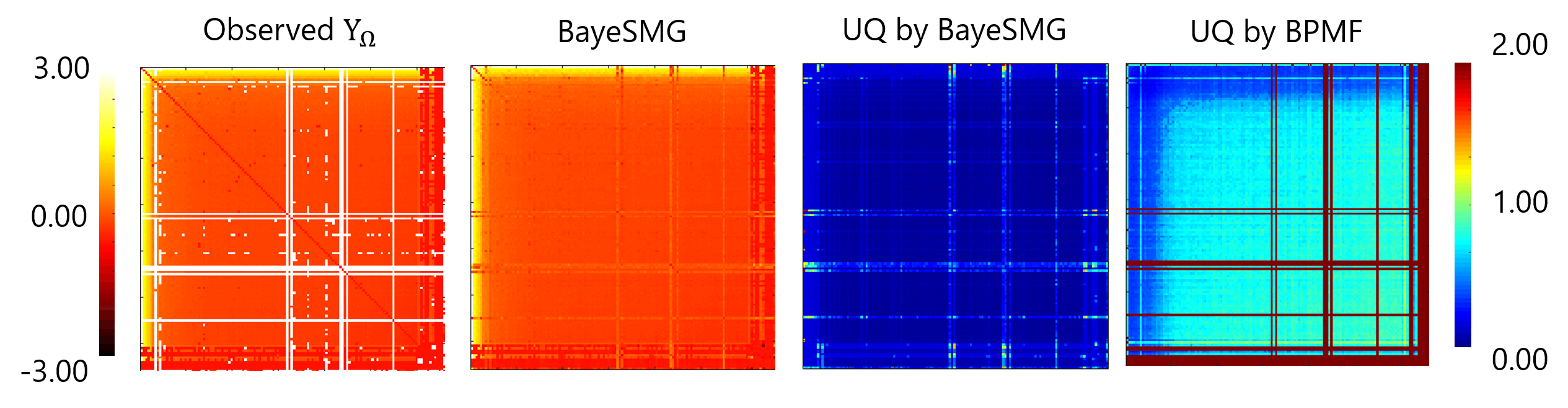}
\vspace{-0.2in}
\caption{Performance comparison between BayeSMG and BPMF on the ambient noise cross-correlation time delay data matrix. The first plot (from the left) shows the observed entries in the delay matrix, with missing entries in white. The second plot shows the completed matrix via the posterior mean from BayeSMG. The third and fourth plots visualize the widths of the entry-wise 95\% HPD intervals from BayeSMG and BPMF.}
\label{fig:sensor}
\vspace{-0.2in}
\end{figure}

One shortcoming of ANSI, however, is that many pairwise cross-correlations do not contain identifiable signals. In other words, the peak in the cross-correlation is unobserved since ANSI works on weak ambient noises. This missing data then results in missing entries in the $133\times 133$ data matrix. To determine whether a cross-correlation is ``missing'', we first identify which correlations have an unsatisfactory signal-to-noise ratio (SNR), by inspecting the standard deviation $\xi$ outside of the main wave lobe relative to the magnitude of the wave peak $g$. The correlation is deemed missing if $g/\xi < 20$. We note that entries on this cross-correlation matrix $\bm{X}$ are observed with noise due to background vibrations caused by bubble collapse and boiling water. Here, the standard deviation of the noise is estimated to be $\eta=0.05$ from an inspection of sensor readings during the period when only noise signals are present; this is then used to initialize $\eta$ in the Gibbs sampler. Figure \ref{fig:sensor} shows the observed noisy matrix entries $\bm{Y}_{\Omega}$.

To proceed with ANSI analysis, one would then need to estimate missing entries in the delay data matrix $\bm{X}$. \cite{bensen2007processing} shows that such a matrix is indeed low-rank. 
Here, uncertainty quantification is crucial for estimating geologic structure and identifying source of activities. With this uncertainty, engineers can better interpret the wave tomography generated from time delay estimates, and identify parts where estimates are accurate and where they are not. This in turn impacts the accuracy of analysis downstream, which subsequently provides greater insight on reconstruction quality.

Figure \ref{fig:sensor} visualizes the recovery and UQ performance from BayeSMG and BPMF, using an estimated rank of $R=15$ via the approach in Section \ref{sec:rank}. We see that the BayeSMG yields much more precise estimates (i.e., narrower HPD interval widths) compared to the BPMF. In particular, when an entire row or column of $\bm{X}$ is missing, the uncertainties returned by BPMF can be very high, which reduces the usefulness of its recovered entries. On the contrary, the proposed BayeSMG method, by leveraging subspace information, can yield more precise inference on these missing rows and columns. One underlying reason is that the BayeSMG approach explicitly integrates subspace modeling for recovery and UQ. From the visualization of $\bm{Y}_{\Omega}$ in Figure \ref{fig:sensor}, we see that there are clearly-seen bright stripes in the left and top edges of the plot, which strongly suggests the presence of low-rank subspaces in $\bm{X}$. It is not a surprise since we know several sensors have highly correlated signals due to their proximity. The BayeSMG appears to exploit this subspace structure to provide more confident predictions. The BPMF yields much higher uncertainty in inference, particularly in rows and columns with little to no observations. While the ground truth for the entire matrix $\bm{X}$ is not known for this sensor network, we would expect from previous experiments that the BayeSMG yields improved recovery performance over the BPMF, particularly in rows and columns with few observations.

With posterior samples on $\bm{X}$ in hand, we can then use its subspace information to locate (or match) a few sensors that contain highly correlated signals with each other. This sensor matching is helpful in seismology studies since we can use it to estimate the dimension and the capacity of the hydrothermal reservoir of the geyser \citep{wu2017anatomy}. We first perform an SVD step on the posterior mean $\hat{\bm{X}}$, and find the singular vector with the largest singular value. We then inspect all the rows of the matrix $\hat{\bm{X}}$, and select the rows most aligned with this vector. We check these rows to locate the most significantly correlated sensors. Figure \ref{fig:geyser}(b) shows the locations of the 12 most correlated sensors and their relative directions from each other. The identified sensors are among the closest to the Old Faithful geyser, and their related observations are dominated by the highly fractured and porous geological structure underground adjacent to the geyser. Using readings from these sensors, researchers can identify a different pattern of the waveform in tremor signals, which suggests a variety of geological structures underneath the geyser. 
\section{Conclusion}
\label{sec:conclusion}
We proposed a new BayeSMG model for uncertainty quantification in low-rank matrix completion. A key novelty of the BayeSMG model is that it parametrizes the unknown matrix $\bm{X}$ via manifold prior distributions on its row and column subspaces. This Bayesian subspace parametrization allows for direct posterior inference on matrix subspaces, which we can use for improved matrix recovery. We introduced a Gibbs sampler for posterior inference, which provides efficient posterior sampling even for matrices with dimensions on the order of thousands. Additionally, we showed that BayeSMG provides a probabilistic interpretation for subspace coherence, which we can use to show an error monotonicity result for UQ. We then showed the effective recovery and UQ performance of BayeSMG on simulated data, image data, and an application for seismic sensor network recovery. Codes for the BayeSMG sampler with illustrative examples will be released in a package in MATLAB.

For future work, it would be interesting to design locations for observations to control the uncertainties, exploring the connection with experimental design literature, e.g., integrated mean-squared error designs \citep{Sea1989} or distance-based designs \citep{MJ2016}. The exploration of this Bayesian uncertainty quantification for guiding sequential sampling of entries (see \citealp{makinformation}) is also of interest. We would also like to investigate further the problem of rank estimation for matrix completion, including theoretical guarantees and an efficient fully Bayesian implementation, extending the work of \cite{hoff2007model}. Another interesting topic to explore is an extension of the i.i.d. Gaussian error assumption to account for skewed or spatially correlated errors.

\section*{Acknowledgments}

Henry Shaowu Yuchi and Yao Xie are supported by NSF CCF-1650913, NSF DMS-1938106, and NSF DMS-1830210. Simon Mak is supported by NSF CSSI Frameworks grant 2004571. The data and picture used in the seismic sensor network recovery are provided by Sin-Mei Wu and Fan-Chi Lin.\\

MATLAB codes for the BayeSMG sampler can be found on \href{https://github.com/henry-gatech/BayeSMG}{GitHub}.

\hfill
\newpage
\bibliography{sample}

\newpage
\appendix
\section{Proofs}
\subsection{Proof of Lemma \ref{thm:smg}}
\begin{proof}
We first prove part (a) of the lemma. To show that $\bm{X} \in \mathcal{T}$ almost surely, let $\bm{Z}$ be an arbitrary matrix in $\mathbb{R}^{m_1 \times m_2}$, with SVD $\bm{Z} = \widetilde{\bm{U}}{\bm{D}}\widetilde{\bm{V}}^T$, ${\bm{D}} = \text{diag}(\{{d}_k\}_{k=1}^R)$. Letting $\bm{u}_k = \mathcal{P}_{\mathcal{U}}\widetilde{\bm{u}}_k$ and $\bm{v}_k = \mathcal{P}_{\mathcal{V}} \widetilde{\bm{v}}_k$, where $\widetilde{\bm{u}}_k$ and $\widetilde{\bm{v}}_k$ are column vectors for $\widetilde{\bm{U}}$ and $\widetilde{\bm{V}}$ respectively, we have $\bm{u}_k \in \mathcal{U}$ and $\bm{v}_k \in \mathcal{V}$ for $k=1, \cdots, R$. From Definition \ref{def:smg}, $\bm{X}$ can then be written as $\bm{X} = \mathcal{P}_{\mathcal{U}}\bm{Z}\mathcal{P}_{\mathcal{V}} = (\mathcal{P}_{\mathcal{U}}\widetilde{\bm{U}}){\bm{D}}(\mathcal{P}_{\mathcal{V}}\widetilde{\bm{V}})^T = \sum_{k=1}^R {d}_k \bm{u}_k \bm{v}_k^T$, as desired. Next, note that the pseudo-inverse of $\mathcal{P}_{\bm{u}}$, $(\mathcal{P}_{\bm{u}})^+$, is simply $\mathcal{P}_{\bm{u}}$, since $\mathcal{P}_{\bm{u}}(\mathcal{P}_{\bm{u}})^+ \mathcal{P}_{\bm{u}} = (\mathcal{P}_{\bm{u}})^+ \mathcal{P}_{\bm{u}} (\mathcal{P}_{\bm{u}})^+ = \mathcal{P}_{\bm{u}}$ by the idempotency of $\mathcal{P}_{\bm{u}}$, and $\mathcal{P}_{\bm{u}}(\mathcal{P}_{\bm{u}})^+ = (\mathcal{P}_{\bm{u}})^+\mathcal{P}_{\bm{u}}$ are both symmetric. Moreover, letting $\det^*$ be the pseudo-determinant operator, we have $\text{det}^*(\mathcal{P}_{\mathcal{U}}) = \text{det}^*(\bm{U}\bm{U}^T) = \det(\bm{U}^T\bm{U}) = 1$, and $\text{det}^*(\mathcal{P}_{\mathcal{V}}) = 1$ by the same argument. Using this along with Theorem 2.2.1 in \cite{GN1999}, the density function $f(\bm{X})$ and the distribution of $\text{vec}(\bm{X})$ follow immediately.

We now prove part (b) of the lemma. From part (a), we have $\textup{vec}(\bm{X}) \sim \mathcal{N}\{\bm{0},\sigma^2 (\mathcal{P}_{\mathcal{V}}\otimes \mathcal{P}_{\mathcal{U}}) \}$, so:
\[ [ \bm{Y}_{\Omega}, \bm{X}_{\Omega^c}] \sim \mathcal{N}\left\{\bm{0},  \begin{bmatrix}
\sigma^2 \bm{R}_N(\Omega) + \eta^2 \bm{I} & \sigma^2 (\mathcal{P}_{\mathcal{V}} \otimes \mathcal{P}_{\mathcal{U}})_{\Omega,\Omega^c} \\
\sigma^2 (\mathcal{P}_{\mathcal{V}} \otimes \mathcal{P}_{\mathcal{U}})^T_{\Omega,\Omega^c} & \sigma^2 (\mathcal{P}_{\mathcal{V}} \otimes \mathcal{P}_{\mathcal{U}})_{\Omega^c}
\end{bmatrix}\right\}.\]
\normalsize The expressions for $\bm{X}^P_{\Omega^c}$ and $\boldsymbol{\Sigma}^P_{\Omega^c}$ in \eqref{eq:condp} then follow from the conditional density of the multivariate Gaussian distribution. Part (c) of the lemma can be shown in a similar way as for part (b).
\end{proof}

\subsection{Proof of Proposition \ref{prop:svd}}
\begin{proof}
\label{proof:svd}
For fixed $\mathcal{P}_{\mathcal{U}}$ and $\mathcal{P}_{\mathcal{V}}$, $\bm{X}$ can be written as:
\begin{equation}
\bm{X} = \mathcal{P}_{\mathcal{U}} \bm{Z} \mathcal{P}_{\mathcal{V}} = \bm{U} (\bm{U}^T \bm{Z} \bm{V}) \bm{V}^T,
\label{eq:svd1}
\end{equation}
where $Z_{i,j} \distas{i.i.d.} \mathcal{N}(0,\sigma^2)$, $\mathcal{P}_{\mathcal{U}} = \bm{U} \bm{U}^T$ and $\mathcal{P}_{\mathcal{V}} = \bm{V} \bm{V}^T$. By Theorem 2.3.10 in \cite{GN1999}, each entry of $\tilde{\bm{Z}} = \bm{U}^T \bm{Z} \bm{V} \in \mathbb{R}^{R \times R}$ follows $\tilde{Z}_{i,j} \distas{i.i.d.} \mathcal{N}(0,\sigma^2)$. Note that the distribution of $\tilde{\bm{Z}}$ is independent of the initial choice of $\mathcal{P}_{\mathcal{U}}$ and $\mathcal{P}_{\mathcal{V}}$ (and thereby $\bm{U}$ and $\bm{V}$). By Theorem 1 of \cite{She2001}, $\tilde{\bm{Z}}$ can be further factorized via its SVD:
\begin{equation}
\tilde{\bm{Z}} = \tilde{\bm{U}} \bm{D} \tilde{\bm{V}}^T,
\label{eq:svd2}
\end{equation}
with independent $\tilde{\bm{U}} \sim U(\mathcal{V}_{R,R})$, $\tilde{\bm{V}} \sim U(\mathcal{V}_{R,R})$ and $\textup{diag}(\bm{D})$ following the repulsed normal distribution  \eqref{eq:ql0}. 

Next, assign independent uniform priors ${U}(\mathcal{G}_{R,m_1-R})$ and ${U}(\mathcal{G}_{R,m_2-R})$ on projection matrices $\mathcal{P}_{\mathcal{U}}$ and $\mathcal{P}_{\mathcal{V}}$, which induces independent uniform priors ${U}(\mathcal{V}_{R,m_1-R})$ and ${U}(\mathcal{V}_{R,m_2-R})$ on frames $\bm{U}$ and $\bm{V}$. From \eqref{eq:svd1}, we have:
\begin{equation}
\bm{X} = \bm{U} (\tilde{\bm{U}} \bm{D} \tilde{\bm{V}}^T) \bm{V}^T = (\bm{U} \tilde{\bm{U}}) \bm{D} (\bm{V} \tilde{\bm{V}} )^T =: \tilde{\tilde{\bm{U}}} \bm{D} \tilde{\tilde{\bm{V}}}^T.
\label{eq:svd3}
\end{equation}
Note that $\tilde{\tilde{\bm{U}}} = \bm{U} \tilde{\bm{U}}$ is an orthonormal frame, since $(\bm{U} \tilde{\bm{U}})^T (\bm{U} \tilde{\bm{U}}) = \tilde{\bm{U}}^T (\bm{U}^T \bm{U}) \tilde{\bm{U}} = \tilde{\bm{U}}^T \tilde{\bm{U}} = \bm{I}$. Moreover, $\tilde{\tilde{\bm{U}}} \sim U(\mathcal{V}_{R,m_1-R})$, since $\bm{U}$ and $\tilde{\bm{U}}$ are independent and uniformly distributed. Similarly, one can show $\tilde{\tilde{\bm{V}}} = \bm{V} \tilde{\bm{V}} \sim U(\mathcal{V}_{R,m_2-R})$ as well, which proves the proposition.
\end{proof}
\subsection{Proof of Lemma \ref{thm:mle}}
\begin{proof}
Since ${U}(\mathcal{G}_{R,m-R})$ is a special case of the matrix Langevin distribution (Section 2.3.2 in \cite{Chi2012}), it follows from (2.3.22) of \cite{Chi2012} that $[\mathcal{P}_{\mathcal{U}}|R] \propto 1$ and $[\mathcal{P}_{\mathcal{V}}|R] \propto 1$. For fixed $\eta^2$ and $\sigma^2$, the MAP estimator for $\bm{X}$ then becomes:
\begin{align*}
\tilde{\bm{X}} &\in \underset{\bm{X} \in \mathbb{R}^{m_1 \times m_2} }{\textup{Argmax}} \; [\bm{Y}_{\Omega}|\bm{X},\eta^2] [\bm{X}|\mathcal{P}_{\mathcal{U}},\mathcal{P}_{\mathcal{V}},\sigma^2,R] \cdot \\
& \hspace{0.12\textwidth} [\mathcal{P}_{\mathcal{U}}|R] \; [\mathcal{P}_{\mathcal{V}}|R] \; [R] \\
& \quad \quad \textup{s.t.} \; \; \mathcal{P}_{\mathcal{U}} \in \mathcal{G}_{R,m_1-R}, \mathcal{P}_{\mathcal{V}} \in \mathcal{G}_{R,m_2-R}, R \leq m_1 \wedge m_2\\
&\in \underset{\bm{X} \in \mathbb{R}^{m_1 \times m_2}}{\textup{Argmax}} \exp\left\{-\frac{1}{2\eta^2} \|\bm{Y}_{\Omega} - \bm{X}_{\Omega}\|_2^2 \right\} \cdot \\
& \hspace{0.075\textwidth}\left[ \frac{1}{(2\pi\sigma^2)^{R^2/2}} \exp\left\{ -\frac{1}{2\sigma^2} \textup{tr}\left[ (\bm{X}\mathcal{P}_{\mathcal{V}})^T (\mathcal{P}_{\mathcal{U}} \bm{X}) \right] \right\} \right] \cdot\\
& \quad \quad \textup{s.t.} \;\; \mathcal{P}_{\mathcal{U}} \in \mathcal{G}_{R,m_1-R}, \mathcal{P}_{\mathcal{V}} \in \mathcal{G}_{R,m_2-R}, R \leq m_1 \wedge m_2\\
&\in \underset{\bm{X} \in \mathbb{R}^{m_1 \times m_2}}{\textup{Argmin}} \left[ \frac{1}{\eta^2} \|\bm{Y}_{\Omega} - \bm{X}_{\Omega}\|_2^2 + \log(2 \pi \sigma^2) R^2 + \right.\\
& \hspace{0.12\textwidth} \left. \frac{1}{\sigma^2}\textup{tr}\left[ (\bm{X}\mathcal{P}_{\mathcal{V}})^T (\mathcal{P}_{\mathcal{U}} \bm{X}) \right] \right]\\
& \quad \quad \textup{s.t.} \;\; \mathcal{P}_{\mathcal{U}} \in \mathcal{G}_{R,m_1-R}, \mathcal{P}_{\mathcal{V}} \in \mathcal{G}_{R,m_2-R}, R \leq m_1 \wedge m_2.
\end{align*}
\normalsize
Since $\bm{X} = \mathcal{P}_{\mathcal{U}} \bm{Z} \mathcal{P}_{\mathcal{V}}$, we have $\bm{X} = \bm{U} \bm{D} \bm{V}^T$ for some $\bm{D} = \text{diag}(\{d_k\}_{k=1}^R)$, $\bm{U} \in \mathbb{R}^{m_1 \times R}$ and $\bm{V} \in \mathbb{R}^{m_2 \times R}$, where $\bm{U}$ and $\bm{V}$ are $R$-frames satisfying $\mathcal{P}_{\mathcal{U}} = \bm{U}\bm{U}^T$ and $\mathcal{P}_{\mathcal{V}} = \bm{V}\bm{V}^T$. Hence:
\begin{align*}
&\textup{tr}\left[ (\bm{X}\mathcal{P}_{\bm{V}})^T (\mathcal{P}_{\bm{U}} \bm{X}) \right]\\
&= \textup{tr}\left[ (\bm{V}\bm{V}^T) (\bm{V} \bm{D} \bm{U}^T) (\bm{U} \bm{U}^T) (\bm{U} \bm{D} \bm{V}^T) \right]\\
&= \textup{tr}\left[ (\bm{V}^T\bm{V})^2 \bm{D} (\bm{U}^T \bm{U})^2 \bm{D} \right] \quad \tag{cyclic invariance of trace}\\
&=  \textup{tr}\left[ \bm{D}^2 \right] \tag{$\bm{V}^T\bm{V} = \bm{I}$ and $\bm{U}^T\bm{U} = \bm{I}$}\\
&= \|\bm{X}\|_F^2, \tag{Frob. norm is equal to Schatten 2-norm}
\end{align*}
which proves the expression in \eqref{eq:matest}.

\end{proof}
\subsection{Proof of Theorem \ref{thm:errred}}
\begin{proof}
\label{proof:gibbs}
Consider the following block decomposition:
\[\bm{R}_{N+1}(\Omega \cup (i,j)) + \gamma^2 \bm{I} = \begin{pmatrix}
\bm{R}_N(\Omega) + \gamma^2 \bm{I} & \boldsymbol{\nu}_i(\mathcal{U}) \circ \boldsymbol{\nu}_j(\mathcal{V})\\
[\boldsymbol{\nu}_i(\mathcal{U}) \circ \boldsymbol{\nu}_j(\mathcal{V})]^T & \mu_i(\mathcal{U})\mu_j(\mathcal{V}) + \gamma^2
\end{pmatrix}.\]
Using the Schur complement identity for matrix inverses \cite{HK1971}, we have:
\begin{equation}
\left[\bm{R}_{N+1}(\Omega \cup (i,j)) + \gamma^2 \bm{I} \right]^{-1} = \begin{pmatrix}
\boldsymbol{\Gamma} + \tau^{-1} \boldsymbol{\Gamma} \boldsymbol{\xi} \boldsymbol{\xi}^T \boldsymbol{\Gamma} & -\tau^{-1} \boldsymbol{\xi}^T \boldsymbol{\Gamma} \\
-\tau^{-1} \boldsymbol{\Gamma} \boldsymbol{\xi} & \tau^{-1}
\end{pmatrix},
\label{eq:blockinv}
\end{equation}
where $\boldsymbol{\xi} = \boldsymbol{\nu}_i(\mathcal{U}) \circ \boldsymbol{\nu}_j(\mathcal{V})$, $\boldsymbol{\Gamma} = \left[\bm{R}_{N}(\Omega) + \gamma^2 \bm{I}\right]^{-1}$ and $\tau = \mu_i(\mathcal{U})\mu_j(\mathcal{V}) - \boldsymbol{\xi}^T\Gamma\boldsymbol{\xi} + \gamma^2$. Using the conditional variance expression in \eqref{eq:varcxij}, $\tau = \text{Var}(X_{i,j}|\bm{Y}_{\Omega})/\sigma^2 + \gamma^2$. Letting $\widetilde{\boldsymbol{\xi}} = \boldsymbol{\nu}_k(\mathcal{U}) \circ \boldsymbol{\nu}_l(\mathcal{V})$ and applying \eqref{eq:varcxij} again, it follows that:
\begin{align*}
&\text{Var}(X_{k,l}|\bm{Y}_{\Omega \cup (i,j)}) \\
&= \sigma^2 \left\{ \mu_k(\mathcal{U})\mu_l(\mathcal{V}) - \widetilde{\boldsymbol{\xi}}^T\Gamma\widetilde{\boldsymbol{\xi}}\right\}\\
& \quad - \tau^{-1} \sigma^2 \left\{ \boldsymbol{\nu}_{i,j}^T \left[ \bm{R}_N(\Omega) + \gamma^2 \bm{I}\right]^{-1} \boldsymbol{\nu}_{k,l} - \nu_{i,k}(\mathcal{U}) \nu_{j,l}(\mathcal{V}) \right\}^2 \tag{using \eqref{eq:blockinv} and algebraic manipulations}\\
&= \text{Var}(X_{k,l}|\bm{Y}_{\Omega}) - \frac{ \text{Cov}^2(X_{i,j},X_{k,l}|\bm{Y}_{\Omega}) }{\text{Var}(X_{i,j}|\bm{Y}_{\Omega}) + \eta^2}, \tag{from \eqref{eq:condp}}
\end{align*}
\normalsize
which proves the theorem.
\end{proof}
\subsection{Proof of Corollary \ref{cor:mono}}
\begin{proof}
This follows directly from Theorem \ref{thm:errred} and the fact that: \[{ \text{Cov}^2(X_{i,j},X_{k,l}|\bm{Y}_{\Omega_{1:N}}) }/\{\text{Var}(X_{i,j}|\bm{Y}_{\Omega_{1:N}}) + \eta^2\} > 0.\]
\end{proof}
\subsection{Proof of full conditional distributions}
\begin{proof}
For fixed rank $R$, the posterior distribution $[\Theta|\bm{Y}]$ can be written as:
\begin{align*}
[\bm{U},\bm{D},\bm{V},\sigma^2|\bm{Y}] & \propto [\bm{Y}|\bm{U}, \bm{D}, \bm{V},\sigma^2] \cdot [\bm{U}] \cdot [\bm{V}] \cdot [\bm{D} |\sigma^2] \cdot [\sigma^2] \\
&  \propto \frac{1}{(\eta^2)^{(m_1 m_2)/2}} \exp\left\{ -\frac{1}{2\eta^2} \|\bm{Y}-\bm{U}\bm{D}\bm{V}^T\|_F^2 \right\} \cdot \frac{1}{(\sigma^2)^{R/2}}\\
& \quad  \cdot \exp\left\{ -\frac{1}{2 \sigma^2} \sum_{k=1}^R d_k^2 \right\} \cdot \prod_{\substack{k,l=1\\ k<l}}^R |d_k^2 - d_l^2| \\
& \quad \cdot \frac{1}{(\sigma^2)^{\alpha_{\sigma^2}+1}} \exp\left\{ -\frac{\beta_{\sigma^2}}{\sigma^2} \right\} \cdot \frac{1}{(\eta^2)^{\alpha_{\eta^2}+1}} \exp\left\{ -\frac{\beta_{\eta^2}}{\eta^2} \right\}.
\end{align*}
\normalsize
From this, the full conditional distributions can then be derived as follows:
\begin{align*}
[\bm{U}|\bm{Y},\bm{D},\bm{V},\sigma^2,\eta^2] & \propto \text{etr}\{ (\bm{Y}\bm{V}\bm{D})^T \bm{U} /\eta^2 \} \sim vMF(m_1,R,\bm{Y}\bm{V}\bm{D}/\eta^2),\\
[\bm{V}|\bm{Y},\bm{U},\bm{D},\sigma^2,\eta^2] & \propto \text{etr}\{ (\bm{Y}^T\bm{U}\bm{D})^T \bm{V}/\eta^2 \} \sim vMF(m_2,R,\bm{Y}^T\bm{U}\bm{D}/\eta^2),\\
[\bm{D}|\bm{Y},\bm{U},\bm{V},\sigma^2,\eta^2] & \propto \exp\left\{ -\frac{1}{2\eta^2} \|\bm{Y}-\bm{U}\bm{D}\bm{V}^T\|_F^2 \right\} \exp\left\{ -\frac{1}{2 \sigma^2} \sum_{k=1}^R d_k^2 \right\} \prod_{\substack{k,l=1\\ k<l}}^R |d_k^2 - d_l^2| \\
& \sim \mathcal{RN}\left( \sigma^2 \text{diag}(\bm{U}^T\bm{Y}\bm{V})/(\eta^2+\sigma^2), \eta^2 \sigma^2 / (\eta^2 + \sigma^2) \right)\\
[\sigma^2|\bm{Y},\bm{U},\bm{D},\bm{V},\eta^2] & \sim IG( \alpha + R/ 2, \beta + \text{tr}(\bm{D}^2) / 2 )\\
[\eta^2|\bm{Y},\bm{U},\bm{D},\bm{V},\sigma^2] & \sim IG( \alpha_{\eta^2} + m_1 m_2 / 2, \beta_{\eta^2} + \|\bm{Y} - \bm{U}\bm{D}\bm{V}^T\|_F^2/2).
\end{align*}
\end{proof}
\end{document}